\documentclass[12pt]{article}   	
\usepackage[margin=1.in] {geometry} 
\usepackage{sidecap}               		
\usepackage{graphicx}				
\usepackage{caption}
\usepackage{adjustbox}		
					
\usepackage{mathtools}
\usepackage{cancel}
\usepackage[toc]{appendix}
\usepackage{dsfont}
\usepackage{multirow}
\usepackage{dblfloatfix}
\usepackage{cleveref}
\usepackage{tikz}
\usepackage{float}
\usepackage{placeins}
\usepackage{authblk}
\usepackage{subcaption}
\usepackage{amsmath}
\usepackage{amssymb}
\usepackage [autostyle, english = american]{csquotes}
\usepackage{scalerel}
\usepackage[numbers]{natbib}
\usepackage{environ}
\usepackage{tikz}

\newcommand\vfrac[2]{\ThisStyle{%
  \setbox0=\hbox{$\SavedStyle#1#2$}%
  \setbox2=\hbox{$\SavedStyle X$}%
  \ifdim\ht0>\ht2\setlength{\ht0}{\ht2}\fi%
  #1\mathord{\stretchto{\raisebox{2.3\LMpt}{$\SavedStyle/$}}{\ht0}}#2}}

\DeclarePairedDelimiter{\abs}{\lvert}{\rvert}
\DeclareMathOperator{\sech}{sech}
\DeclareMathOperator{\erfc}{erfc}

\begin{document}
\date{\today}
\title{
Plasma electron hole oscillatory velocity instability
}

\author[1]{Chuteng Zhou}
\author[2]{Ian H. Hutchinson}
\affil[1,2]{Plasma Science and Fusion Center, Massachusetts Institute of Technology, Cambridge, MA, USA}
\date{}
\maketitle

\begin{abstract}
In this paper, we report a new type of instability of electron holes (EHs) interacting with passing ions. The nonlinear interaction of EHs and ions is investigated using a new theory of hole kinematics. It is shown that the oscillation in the velocity of the EH parallel to the magnetic field direction becomes unstable when the hole velocity in the ion frame is slower than a few times the cold ion sound speed. This instability leads to the emission of ion-acoustic waves from the solitary hole and decay in its magnitude. The instability mechanism can drive significant perturbations in the ion density. The instability threshold, oscillation frequency and instability growth rate derived from the theory yield quantitative agreement with the observations from a novel high-fidelity hole-tracking Particle-In-Cell (PIC) code.
\end{abstract}

\section{Introduction}
A plasma electron hole is a solitary vortex structure of trapped electrons in the electron phase-space. The deficit of electrons on the trapped orbits gives rise to a coherent self-trapping structure. We have performed 1-D Particle-In-Cell simulations of EHs with different velocities relative to the ions. Oscillations in the velocity of a slowly moving EH are observed to be unstable. Using a novel theory of hole kinematics that treats the EH as a holistic object, we give a physical explanation for this newly observed instability. The analytical results agree quantitatively with our simulation observations. EHs occur widely in nature and they are important to many space physics phenomena.

The EH was first discovered as a nonlinear equilibrium of a collisionless Vlasov-Poisson plasma under the name of Bernstein-Greene-Kruskal mode \cite{Bernstein1957a}. The study of these nonlinear structures in plasma gained increasing interest after space probe measurements confirmed their wide-spread existence in the Earth's auroral zone \cite{Ergun1998}, magnetosphere \citep{Pickett2004} and in the solar wind \cite{Malaspina2013}. They are implicated in magnetic reconnection \cite{Drake2003a}, electron acceleration \cite{Mozer2016}, collisionless shocks \cite{Wilson2010} and other important plasma dynamics in space. EHs are also detected in the laboratory plasma during magnetic reconnection \citep{Fox2008} and beam injection \citep{Lefebvre2010}.

The early theoretical research of EHs neglected the ion dynamics for simplicity and considered them as a uniform neutralizing background \cite{Schamel1972}. Later, Saeki et al. \citep{Saeki1998} showed using PIC simulations that an EH can be disrupted by ions when its velocity is slower than the ion sound speed. Eliasson et al. \cite{Eliasson2004a} reported that a standing EH can be ejected by the ion density cavity it created and is attracted to ion density maxima. The recent observations of ``slow" EHs also suggest a more important role for the ions. EHs traveling with the ion sound speed $c_s$ have been recently reported at a magnetic reconnection site \cite{Khotyaintsev2010} and the magnetopause \cite{Norgren2015} measured by the Cluster spacecrafts, a velocity much slower than what was frequently observed before ($\sim v_{th,e}$). The authors suggested that Buneman instability resulting from dynamic reconnecting current sheets generated these slow EHs. Schamel gave an upper limit \cite{Schamel1986} for the speed of the EHs by a structural existence argument. How slow can an EH travel? Saeki et al. \citep{Saeki1998} briefly touched upon this question by deriving the nonlinear dispersion relation using the Sagdeev pseudo-potential method. However, the nonlinear dispersion relation is based on the existence of a stationary solution of which the \emph{stability} is not guaranteed. An EH can experience different kinds of instabilities in higher dimensions, e.g.\ the whistler instability \citep{Newman2001} in the strongly magnetized case and the transverse instability \citep{Muschietti2000} in the weakly magnetized case. These instability mechanisms don't involve ion dynamics. Ion-acoustic wave radiation from a solitary structure in plasma has been studied in the case of Langmuir soliton \citep{Schamel1998} and ion hole \citep{Luque2005}. Dyrud et al. \citep{Dyrud2006} reported the observation of ion-acoustic waves emitted from a chain of electron holes in PIC simulations. Dokgo et al. \citep{Dokgo2016} reported the generation of coherent ion-acoustic solitary waves from an EH as it propagates from a lower plasma density region to a higher plasma density region. In this paper, we show that the ion dynamics is important for the stability of an EH even in the 1-D equilibrium, causing an oscillatory velocity instability for slow electron holes to decay into ion-acoustic waves. This discovery suggests that the solitary solutions constructed using the Sagdeev pseudo-potential method can be unstable to small perturbations in its velocity when the ion dynamics becomes important.

The instability mechanism discussed in this paper is closely related to the velocity thus the kinematics of an EH. The EH kinematics has been studied by the authors \citep{Hutchinson2016a} treating the EH as a composite object. When an EH accelerates, ion and electron plasma momentum changes both inside and outside the EH. The EH should move in a way to conserve the total plasma momentum. This quasi-particle approach using the global momentum conservation can be found in the early EH theory developed by Dupree et al. \citep{Dupree1983a} Our theory is a major improvement over Dupree's and was successful in explaining quantitatively the dynamics of EHs observed in PIC simulations \citep{Zhou2016}, such as the transient self-acceleration and the ``hole pushing/pulling" effect due to steady-state hole momentum coupling to the ions. In this paper, we extend our theory to the frequency domain and use multiple-scale analysis to give a mathematically rigorous treatment of the instability.

The paper is organized as follows: in Section \ref{sec:1}, we report the observational details of this instability from our PIC simulation. In Section \ref{sec:2}, a first principle analytic calculation using hole kinematics theory is presented, the instability boundary, unstable mode frequency and growth rate are analytically derived and compared with the PIC simulation observation. In Section \ref{sec:3}, we are going to discuss the nonlinear stage of the instability and its potential implication in space plasma. Section \ref{sec:4} is the summary.

\section{PIC observation of the instability} \label{sec:1}

The simulations are performed using a 1-D electrostatic PIC code with fully kinetic ions, which is designed to study highly-resolved EH dynamics. A solitary EH is created in our simulation using an electron phase-space density deficit as the initial seed \citep{Zhou2016}. The thermal noise in our PIC simulation is controlled by using more than $10^6$ particles per cell. There are $\sim 10$ cells per Debye length to resolve the detail of an EH. We performed box simulation of a solitary EH with a computation domain that can self-consistently follow the EH motion \citep{Zhou2016}. The size of the computation domain is only $\sim 50\lambda_{\mathrm{De}}$ by virtue of using hole tracking. Our hole-tracking PIC allows us to study the detail of hole motion with reasonable computational cost. Once a steady-state EH is obtained in our simulation, we apply an artificial slow ion acceleration to slowly ``push" the EH with ions so that it slows down in the ion frame. This ``pushing" process has been demonstrated to be quasi steady-state and reversible \citep{Zhou2016}. We discovered when performing these ``pushing" runs that there is a limit velocity of the EH in the ion frame, below which the EH becomes unstable. This threshold velocity is well above other physical limit of the system such as the ion reflection velocity limit.

\begin{figure}[htb]
\centering
\includegraphics[width=0.7\textwidth]
{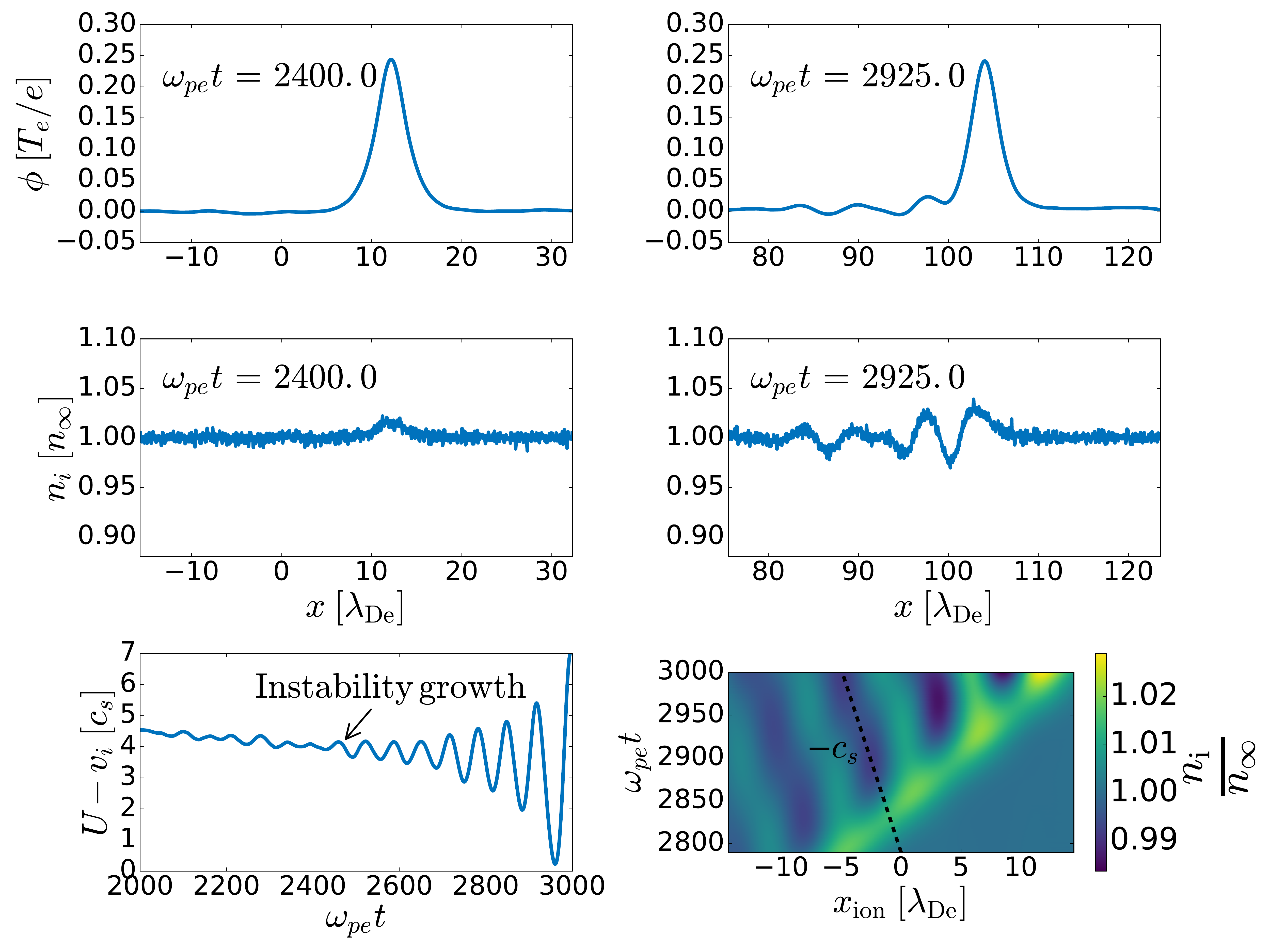}
\caption{The hole potential (first row) and the ion density (second row) before (left) and after (right) the instability growth. Bottom left panel shows the EH velocity in the ion frame and the bottom right panel shows the ion density perturbations due to the EH and the instability. The bulk electrons are Maxwellian at rest in the lab frame and $T_e/T_i=20$.}
\label{instability_example}
\end{figure}

An example of the instability observed in our PIC simulation is presented in Figure \ref{instability_example}. On the top left, we show the characteristic solitary potential structure of a stable EH that extends over several Debye lengths. In a steady-state EH, the ions are slowed down by the hole potential and their density is slightly higher inside the EH. This ion density compressional pulse is the ion-acoustic soliton attached to the phase-space EH described by Saeki et al.\citep{Saeki1998} It is clearly visible inside the stable EH shown on the left. When the EH slows down in the ion frame, the oscillation amplitude in its velocity begins to grow once its speed is slower than a threshold value. The EH potential and the ion density at a later time step ($\omega_{pe}t=2925$) after the instability has grown are presented on the right for comparison. The EH keeps its potential shape while its velocity oscillates. The down stream ion density becomes unsteady as the velocity oscillation amplitude grows. Unsteady ion density perturbations are emitted from the EH after the instability onset. The perturbations propagate in the ion frame with the ion sound speed, mainly in the opposite direction to the EH velocity. The EH velocity was obtained from the hole-tracking module in our code \citep{Zhou2016} and a low pass filter has been applied to it to filter out the statistical noise. The ``hole pushing" can be turned off at any moment before the instability onset and the EH will enter a stable steady-state with the same velocity, but not after the EH is slower than the threshold velocity.

\begin{figure}[hbt]
\centering
\includegraphics[width=0.7\textwidth]
{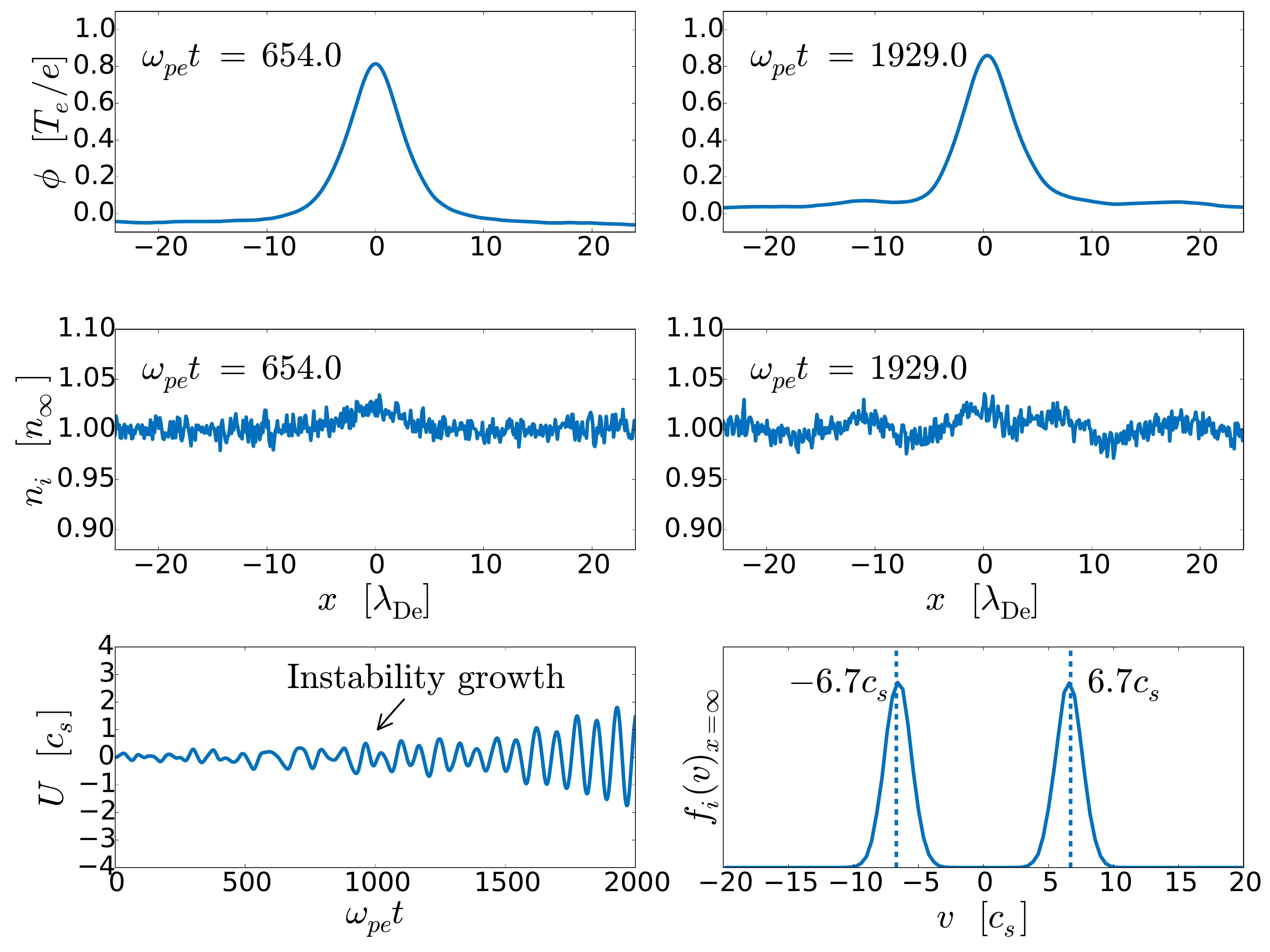}
\caption{The hole potential (first row) and the ion density (second row) before (left) and after (right) the instability growth in a plasma with counter-streaming ions. Bottom left panel shows the EH velocity and the bottom right panel shows the ion distribution function with counter-streaming Maxwellians. The ion streams have an average velocity of $\pm 6.7c_s$ and $T_i=T_e$. The bulk electrons are Maxwellian at rest in the lab frame.}
\label{instability_CounterStreaming}
\end{figure}

A similar phenomenon also happens in a plasma with counter-streaming ions. We initialize an EH at rest ($U=0$ initially) in the lab frame on top of the electron distribution with counter-streaming ions traveling at $\pm v_i$ in the lab frame. This time we don't need to apply any special technique such as the hole pushing and hole tracking. The initialization will naturally favor the formed EH to stay at $U=0$ and we can do a regular box simulation of the EH with a static domain. We observe that there is a minimum value of $v_i$ below which the system is unstable. A case of the observed instability is shown in Figure \ref{instability_CounterStreaming}. The self-consistently formed EH is unstable. Perturbations in its velocity grow exponentially. Ion-acoustic perturbations grow and are emitted from both sides of the EH because the ions are counter-streaming. The simulation was performed with warm ions $T_i=T_e$, corresponding to a case where ion-acoustic waves are strongly Landau damped, ion-ion type of instability and Buneman instability are ruled out by the simulation parameter setting. For this particular case shown in Figure \ref{instability_CounterStreaming}, using a slightly higher $v_i=7c_s$ can stabilize the instability. We shall see later in the paper that from the stability point of view, the counter-streaming ion case is equivalent to the single ion stream case.

We have repeated the numerical experiments with different parameter settings. It is observed that for the same EH, the instability only depends significantly on the relative velocity between the EH and the ions $U-v_i$. The EH velocity with respect to the bulk electrons $U$ and the relative velocity drift between the ions and the bulk electrons $v_i$ alone have no significant influence on the instability onset and its growth. We have performed simulations with an electron-to-ion temperature ratio $T_e/T_i$ from $20$ down to $0.5$. We observe that this velocity instability clearly persists in the regime $T_e\leq T_i$, where ion-acoustic type of instability is unexpected. A hotter ion population leads to a higher threshold value of $U-v_i$ and damped ion-acoustic wings. We shall discuss the finite ion temperature effect on the instability in detail later in our paper.

Our PIC simulation contains no imprecision beyond the statistical noise which was largely mitigated by using a large number of particles. It is clear that a self-consistent solitary solution with a complete ion response can be constructed in the case of instability using the Sagadeev pseudo potential or the BGK approach. Our PIC code actually does this by solving Poisson's equation numerically. However, once this steady-state solitary solution is allowed to evolve in time, it becomes unstable. The characteristics of this instability does not fit into any existing linear plasma stability theory. The core of this problem is very nonlinear because of the strong particle trapping nonlinearity in the EH. We need to adopt a new approach to analyze its mechanism.

\section{Hole velocity stability deduced from kinematics} \label{sec:2}
\subsection{Frequency response of the momentum rate of change}
\begin{figure}[hbt]
\centering
\includegraphics[width=0.4\textwidth]{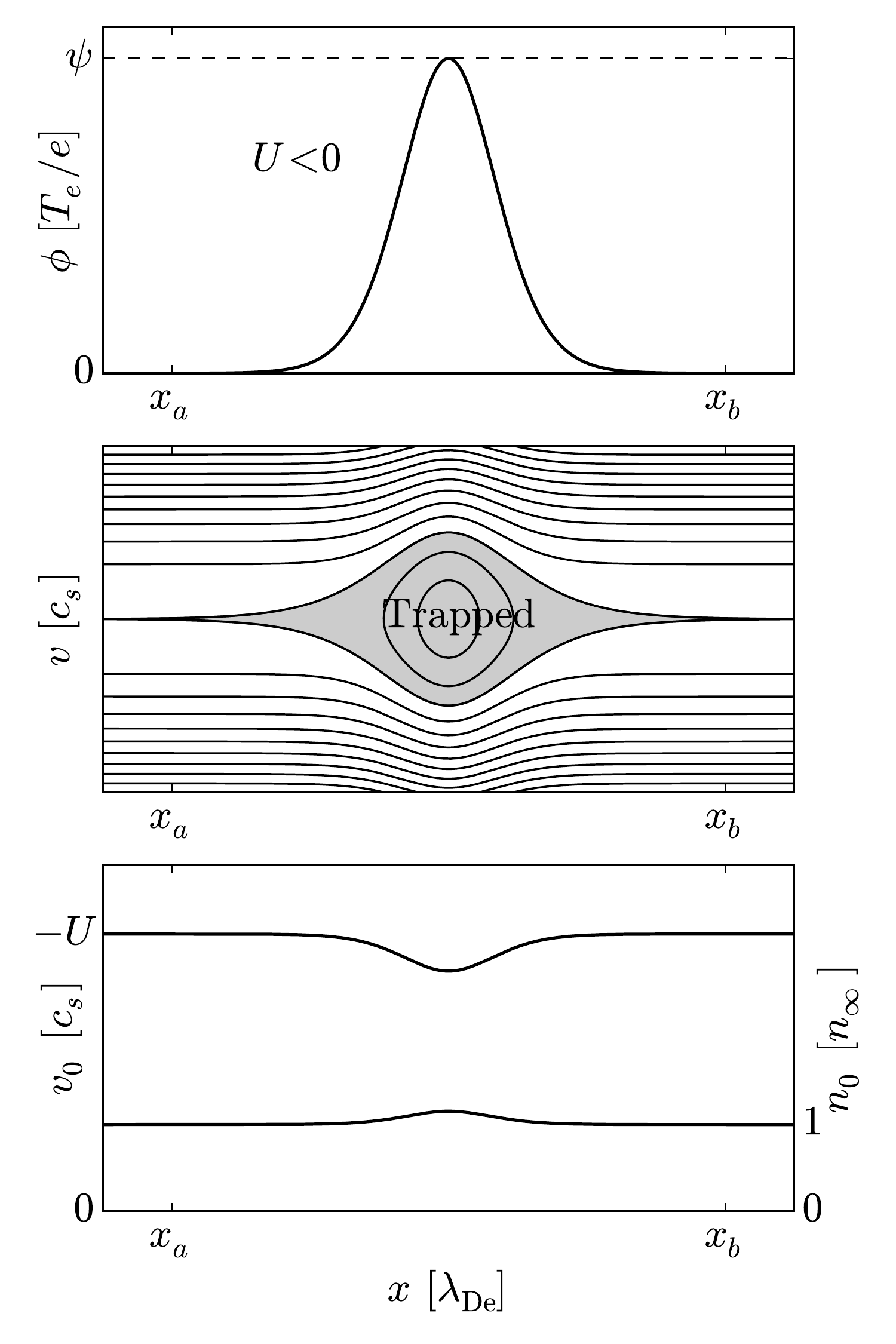}
\caption{Schematic of a steady-state EH with the associated phase-space structure and the ion response. Top: EH potential, middle: electron phase space orbits, the trapped orbits are shaded, bottom: the steady-state ion velocity $v_0$ and density $n_0$ in the hole frame.}
\label{EH_schematic}
\end{figure}

To analyze this instability, we first consider the steady-state solution of an EH. The steady-state EH potential $\phi(x)$ is considered to extend from $x_a$ to $x_b$ in the hole frame, $x_a$ and $x_b$ are taken to be far away from the center of the hole so that both $\phi(x)$ and its derivatives vanish at these limits: $\phi(x_a)=\phi(x_b)=\phi'(x_a)=\phi'(x_b)=0$. The ions and the bulk electrons are assumed to be Maxwellian at rest in the lab frame with their background density being denoted by $n_{\infty}$. The EH moves at a velocity $U$ in the lab frame. The sign convention is such that $U<0$. We first adopt a cold beam approximation for ions and the finite ion temperature effect will be treated later on in this paper. The distance is normalized to $\lambda_{\mathrm{De}}$, $\phi$ is measured in $T_e/e$, the velocity is in units of $c_s=\sqrt{T_e/m_i}$ and the time is normalized to $1/\omega_{pi}$. The schematic of a steady-state EH is shown in Figure \ref{EH_schematic}. The steady-state velocity and density of the ions in the hole frame can be derived from conservation of energy and the continuity equation:
\begin{equation}\label{eqn:equilib}
\left \{ \begin{aligned}
&v_0(x)\,=\,-\frac{U}{|U|}\sqrt{U^2-2\phi(x)}\\
&n_0(x)\,=\,n_{\infty}\frac{|U|}{\sqrt{U^2-2\phi(x)}}
\end{aligned}\right.
\end{equation}

In our previous paper \cite{Hutchinson2016a}, we have established that the motion of an EH is governed by the momentum conservation when acceleration or growth are steady. The parallel momentum contained in the electromagnetic field can be ignored for a field-aligned solitary electrostatic structure. The momentum balance is between the two components of the plasma: $\dot{P}_i+\dot{P}_e=0$, where $\dot{P}_{i/e}$ represents the total inertial-frame momentum rate of change for the two species in a control volume that extends from $x_a$ to $x_b$. Here we extend the analysis to oscillatory acceleration or the frequency domain. To a first approximation, we are going to assume that the EH potential $\phi(x)$ in the hole frame does not change and there is a small perturbation in the hole velocity represented by an ansatz $\dot{U}\sim\mathrm{Re} (\exp[-i\omega t])$. The frequency $\omega$ we consider is much lower than the average passing electron transit frequency. The passing electrons feel a nearly constant hole acceleration during their transit (this amounts to a short-transit-time approximation for electrons), so that we can use the previous expression \cite{Hutchinson2016a} of $\dot{P}_e$ for $U\ll v_{th,e}$ with a full ion response correction \cite{Zhou2016} 
\begin{equation} \label{eqn:electron_momentum}
\dot{P}_e\,=\,-m_e\dot{U}n_\infty\int_{x_a}^{x_b}h(\sqrt{\phi(x)})+1-\frac{\abs{U}}{\sqrt{U^2-2\phi(x)}}\,dx,
\end{equation}
where $h(\chi)\,=\,-\frac{2}{\sqrt{\pi}}\chi+(2\chi^2-1)e^{\chi^2}\erfc(\chi)+1$. The ion response correction accounts for the ion density accumulation inside the EH. The exact shape of the trapped electron distribution doesn't appear in our approach. The total number of trapped electrons inside the EH is deduced from global charge neutrality of the solitary structure. Our ansatz treats the trapped electron phase-space structure as a holistic object.

The ions however feel an oscillating potential when they transit the hole region. The ion momentum change can be decomposed into two different terms, a momentum outflow term $\dot{P}_{io}$ at the boundaries and a contained momentum term $\dot{P}_{ic}$. The conservation of momentum needs to be evaluated at a fixed time. Let subscripts $s$ and $f$ refer to the starting time and the final time, $a$ and $b$ refer to the positions $x_a$ and $x_b$ in the hole frame and bar denote velocities in the inertial frame (the unbarred velocities are evaluated in the hole frame). An ion particle enters the control volume at $x_a$ when $t=t_s$ exits at $x_b$ when $t=t_f$. At $t=t_f$, we have therefore
\begin{eqnarray} \label{eqn:outflow}
\begin{aligned}
\frac{\dot{P}_{io}}{m_i}\,&=\,n_{bf}v_{bf}\bar{v}_{bf}-n_{af}v_{af}\bar{v}_{af} \\ 
\,&=\,n_{bf}v_{bf}\bar{v}_{bf}-n_{bf}v_{bf}\bar{v}_{af}+n_{bf}v_{bf}\bar{v}_{af}-n_{af}v_{af}\bar{v}_{af}\\
\,&=\,n_{bf}v_{bf}(\bar{v}_{bf}-\bar{v}_{af})+(n_{bf}v_{bf}-n_{af}v_{af})\bar{v}_{af}.
\end{aligned}
\end{eqnarray}
The term $\bar{v}_{bf}-\bar{v}_{af}$ represents the ``jetting" effect \cite{Hutchinson2016a} due to the hole acceleration. In the comoving frame of the EH, the equation of motion of a single ion particle admits a first integral
\begin{equation} \label{eqn:conservation_law}
\frac{1}{2}v^2+\phi(x)+\int \dot{U}v\,dt\,=\,\mathrm{Constant}.
\end{equation}
Applying this conservation law between the time $t_s$ and $t_f$, we have
\begin{equation} \label{eqn:conservation}
v_{bf}^2-v_{as}^2+2\int_{t_s}^{t_f}\dot{U}v\,dt=0.
\end{equation}
The equilibrium velocity for $\dot{U}=0$ is $v_{0}(x)=-\frac{U}{\abs{U}}\sqrt{U^2-2\phi(x)}$. The idea is to do an expansion around the equilibrium orbit. Perturbation expansion gives $v=v_0+v_1$ and $\abs{v_1}/\abs{v_0}\sim \abs{t_{ab}\dot{U}/U}\ll 1$ where $t_{ab}=t_f-t_s$ is the single ion transit time. In principle, the amplitude of the hole acceleration can be made arbitrarily small to satisfy this ordering. To the leading order in the small parameter $t_{ab}\dot{U}/U$, we expand the difference between the ion velocity exiting and entering the hole region $v_{bf}-v_{as}$
\begin{equation}\label{eqn:jetting_term}
\begin{aligned}
v_{bf}-v_{as}\,&=\,\frac{-2}{v_{bf}+v_{as}}\int_{t_s}^{t_f}\dot{U}v\,dt\\
&\simeq\,\int_{x_a}^{x_b}\frac{\dot{U}(t(x,x_b))}{U}\,dx.
\end{aligned}
\end{equation}
where $t(x,x_b)=t_f-\int_{x}^{x_b}\frac{du}{v(u)}$ is an intermediate time. To keep notations simple, we will omit its explicit form while keeping in mind that unless stated otherwise, $\dot{U}$ and $v$ are evaluated when the considered ion particle is at the position indicated by the dummy variable of the integration. Taking into account the change in the hole velocity between $t_s$ and $t_f$, we have
\begin{equation}
\begin{aligned}
\bar{v}_{bf}-\bar{v}_{af}\,&=\,v_{bf}-v_{as}+\int_{t_s}^{t_f}\dot{U}\,dt\\
&\simeq\,\int_{x_a}^{x_b}(\frac{1}{U}+\frac{1}{v_0})\dot{U}\,dx.
\end{aligned}
\end{equation}
Thus we have obtained the ``jetting" effect due to the acceleration of the EH to the relevant order and the first term in Eqn. \eqref{eqn:outflow} can be evaluated as
\begin{equation}\label{eqn:outflow_part1}
n_{bf}v_{bf}(\bar{v}_{bf}-\bar{v}_{af})\simeq n_{\infty}\int_{x_a}^{x_b}(-1-\frac{U}{v_0})\dot{U}\,dx.
\end{equation}
To calculate the second term in Eqn. \eqref{eqn:outflow}, we need to know how the ion flux changes with $\dot{U}$. We apply the continuity of an ion fluid element from $x_a$ to $x_b$
\begin{equation}
n_{as}v_{as}\delta t_{as}=n_{bf}v_{bf}\delta t_{bf},
\end{equation}
where $\delta t_{as}$ and $\delta t_{bf}$ are two infinitesimal time durations for the same ion fluid element to enter and exit the control volume. They are related by the derivative of transit time $t_{ab}$ w.r.t.\ the starting time $t_s$: $\delta t_{bf}\simeq\delta t_{as}(1+\frac{d t_{ab}}{dt_s})$. Thus, to the leading order,
\begin{eqnarray} \label{eqn:second_term}
\begin{aligned}
n_{bf}v_{bf}-n_{af}v_{af}&\,=\,n_{bf}v_{bf}-n_{bf}v_{bf}\frac{\delta t_{bf}}{\delta t_{as}}\frac{v_{af}}{v_{as}}\\
&\,=\,n_{bf}v_{bf}\bigg[1-\frac{\delta t_{bf}}{\delta t_{as}}(1+\frac{v_{af}-v_{as}}{v_{as}})\bigg] \\
&\,\simeq n_{bf}v_{bf}(-\frac{dt_{ab}}{dt_s}-\frac{v_{af}-v_{as}}{v_{as}}),
\end{aligned}
\end{eqnarray}
where we used the constancy of inflow density $n_{as}=n_{af}=n_\infty$ and that $dt_{ab}/dt_s$ is of the same order as $(v_{af}-v_{as})/v_{as}$. The derivative $dt_{ab}/dt_s$ describes the non-constancy of the ion transit time due to the hole acceleration. It can be evaluated to the first order using $t_{ab}=\int_{x_a}^{x_b}\frac{dx}{v}$ and the conservation law of Eqn. \eqref{eqn:conservation_law}
\begin{equation} \label{eqn:t_ab}
\begin{aligned}
\frac{dt_{ab}}{dt_s}\,&=\,\frac{d}{dt_s}\int_{x_a}^{x_b}\frac{1}{v}\,dx\\
&=\,\int_{x_a}^{x_b}-\frac{1}{v^3}\frac{\partial(v^2/2)}{\partial t_s}\bigg|_x\,dx\\
&=\,\int_{x_a}^{x_b}-\frac{1}{v^3}\frac{\partial}{\partial t_s}\bigg|_x\left(v_{as}^2/2-\phi(x)-\int_{x_a}^{x}\dot{U}\,dx_1\right)\,dx\\
&\simeq\,\int_{x_a}^{x_b}\frac{1}{v^3}[-U\dot{U}(t_s)+\int_{x_a}^{x}\ddot{U}\,dx_1]\,dx.
\end{aligned}
\end{equation}
$\ddot{U}$ is the rate of change of hole acceleration or the jerk evaluated when the ion particle is at $x_1$. We used interchangeably $\partial/\partial t_s$ and $\partial/\partial t$ as $dt=dt_s(1+\mathcal{O}(t_{ab}\dot{U}/U))$ for $x_1$ fixed. We can further get rid of the $\dot{U}(t_s)$ term in Eqn. \eqref{eqn:t_ab} performing integration by parts
\begin{equation}\label{eqn:integration_by_parts}
\begin{aligned}
\dot{U}(t_s)-\frac{1}{U}\int_{x_a}^{x}\ddot{U}\,dx_1\,&=\,\dot{U}(t_s)-\frac{1}{U}\int_{x_a}^{x}v\,d\dot{U}\\
&\simeq\,\dot{U}(t_s)-\bigg[\frac{\dot{U}(t)v_0(x)}{U}-\frac{\dot{U}(t_s)v_0(x_a)}{U}\bigg]+\frac{1}{U}\int_{x_a}^{x}\dot{U}\frac{dv_0}{dx_1}\,dx_1\\
&\simeq\,-\frac{\dot{U}(t)v_0(x)}{U}-\frac{1}{U}\int_{x_a}^x\frac{\dot{U}\phi'(x_1)}{v_0(x_1)}\,dx_1.
\end{aligned}
\end{equation}
We used here $dv_0/dx_1=-\phi'/v_0$, where $\phi'$ is the spatial derivative of $\phi$. Combining Eqns. \eqref{eqn:t_ab} and \eqref{eqn:integration_by_parts}, we can evaluate the right hand side of Eqn. \eqref{eqn:second_term} as
\begin{equation}\label{eqn:outflow_part2}
\begin{aligned}
n_{bf}v_{bf}(-\frac{dt_{ab}}{dt_s}-\frac{v_{af}-v_{as}}{v_{as}})\,&\simeq\,n_{\infty}(-U)\int_{x_a}^{x_b}-\frac{1}{v_0^3}\bigg[-U\dot{U}(t_s)+\int_{x_a}^x\ddot{U}\,dx_1\bigg]-\frac{1}{U}\frac{\dot{U}}{v_0}\,dx\\
&\simeq\,n_{\infty}\int_{x_a}^{x_b}\frac{U^2}{v_0^3}\bigg[\frac{\dot{U}v_0}{U}+\frac{1}{U}\int_{x_a}^x\frac{\dot{U}\phi'}{v_0}\,dx_1\bigg]+\frac{\dot{U}}{v_0}\,dx\\
&=\,n_{\infty}\int_{x_a}^{x_b}\frac{U}{v_0^2}\dot{U}+\frac{U}{v_0^3}\bigg(\int_{x_a}^x\frac{\dot{U}\phi'}{v_0}\,dx_1\bigg)+\frac{\dot{U}}{v_0}\,dx.
\end{aligned}
\end{equation}
We choose the inertial frame as the instantaneous rest frame of the EH so that $\bar{v}_{af}=v_{af}\simeq-U$, we have a final expression for the total rate of momentum outflow by using Eqns. \eqref{eqn:outflow_part1} and \eqref{eqn:outflow_part2}. It is first order in $t_{ab}\dot{U}/U$, thus linear in $\dot{U}$
\begin{equation} \label{eqn:io}
\dot{P}_{io}=m_i n_\infty \int_{x_a}^{x_b}(-1-2\frac{U}{v_0}-\frac{U^2}{v_0^2})\dot{U}-\frac{U^2}{v_0^3}\bigg(\int_{x_a}^{x}\frac{\dot{U}\phi'}{v_0}\,dx_1\bigg)\,dx.
\end{equation}
Now we proceed to calculate the rate of change of ion momentum contained inside the control volume between $x_a$ and $x_b$ in the same inertial frame at $t=t_f$
\begin{equation} 
\dot{P}_{ic}=m_i\int_{x_a}^{x_b}\frac{\partial (nv)}{\partial t} \,dx+m_i\dot{U}\int_{x_a}^{x_b}n\,dx.
\end{equation}
The derivation of $\dot{P}_{ic}$ is similar in spirit to what we have shown for $\dot{P}_{io}$ but involves heavier algebra, we leave it to the Appendix A. The final result which is the same order as $\dot{P}_{io}$ in Eqn. \eqref{eqn:io} can be expressed as
\begin{equation} \label{eqn:contained}
\dot{P}_{ic}=-m_in_{\infty}\int_{x_a}^{x_b}\int_{x_a}^{x}\frac{U}{v_0^3}\int_{x_a}^{x_1}\dot{U}(t(x_2,x))\phi''(x_2)\,dx_2\,dx_1\,dx.
\end{equation}
We combine Eqs. \eqref{eqn:io} and \eqref{eqn:contained} to give a full expression for $\dot{P}_i=\dot{P}_{io}+\dot{P}_{ic}$. The conservation of total momentum gives an eigenmode equation for $\omega$: 
\begin{equation}\label{eqn:eigenmode}
\dot{P}_i(\omega)+\dot{P}_e(\omega)=0. 
\end{equation}
The imaginary part of $\omega$ determines the stability of the corresponding eigenmode.We apply Nyquist stability analysis \cite{Nyquist1932} to determine the stability. 

\begin{figure}[t]
\centering
\includegraphics[width=0.43\textwidth]
{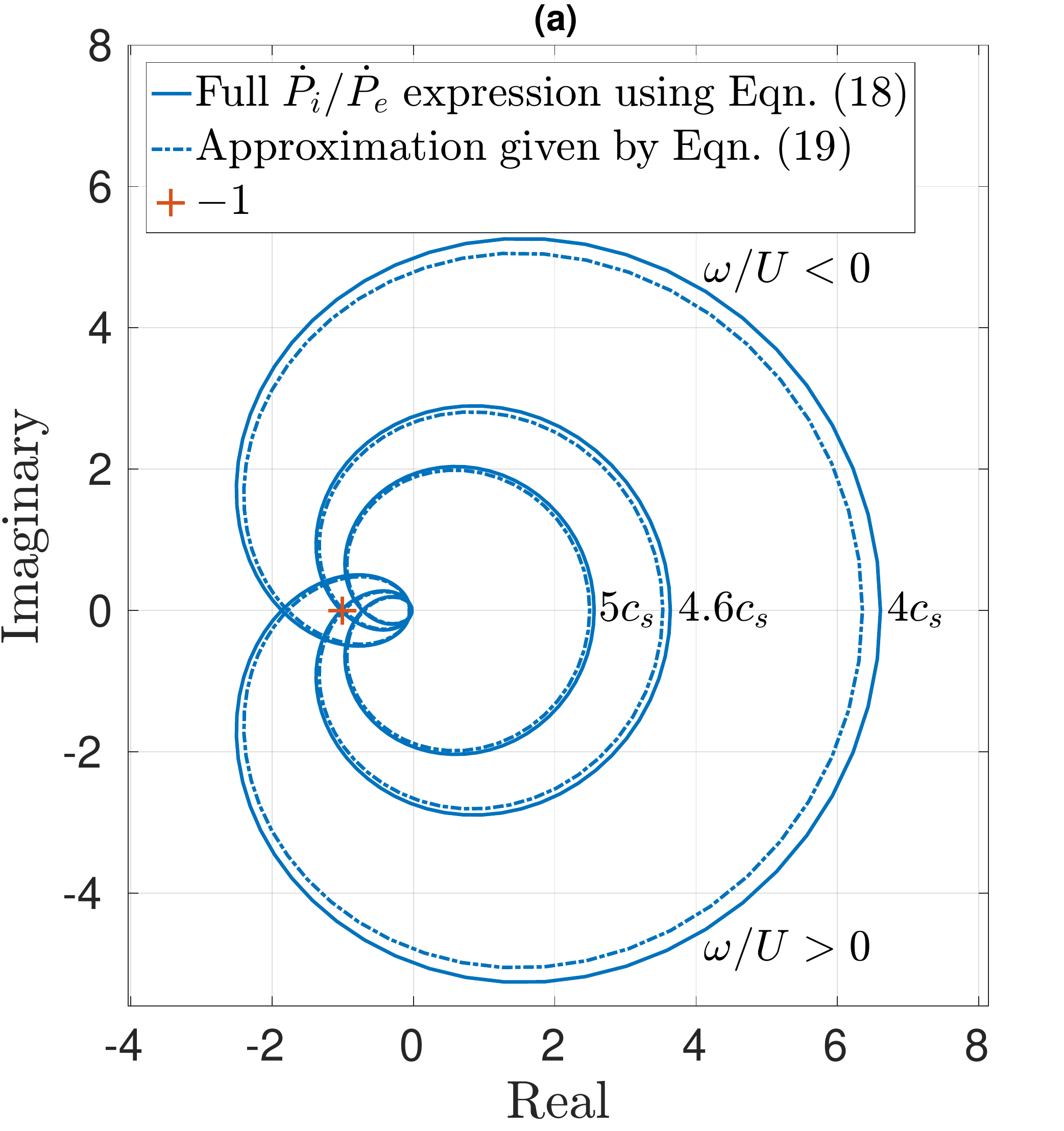}
\includegraphics[width=0.43\textwidth]
{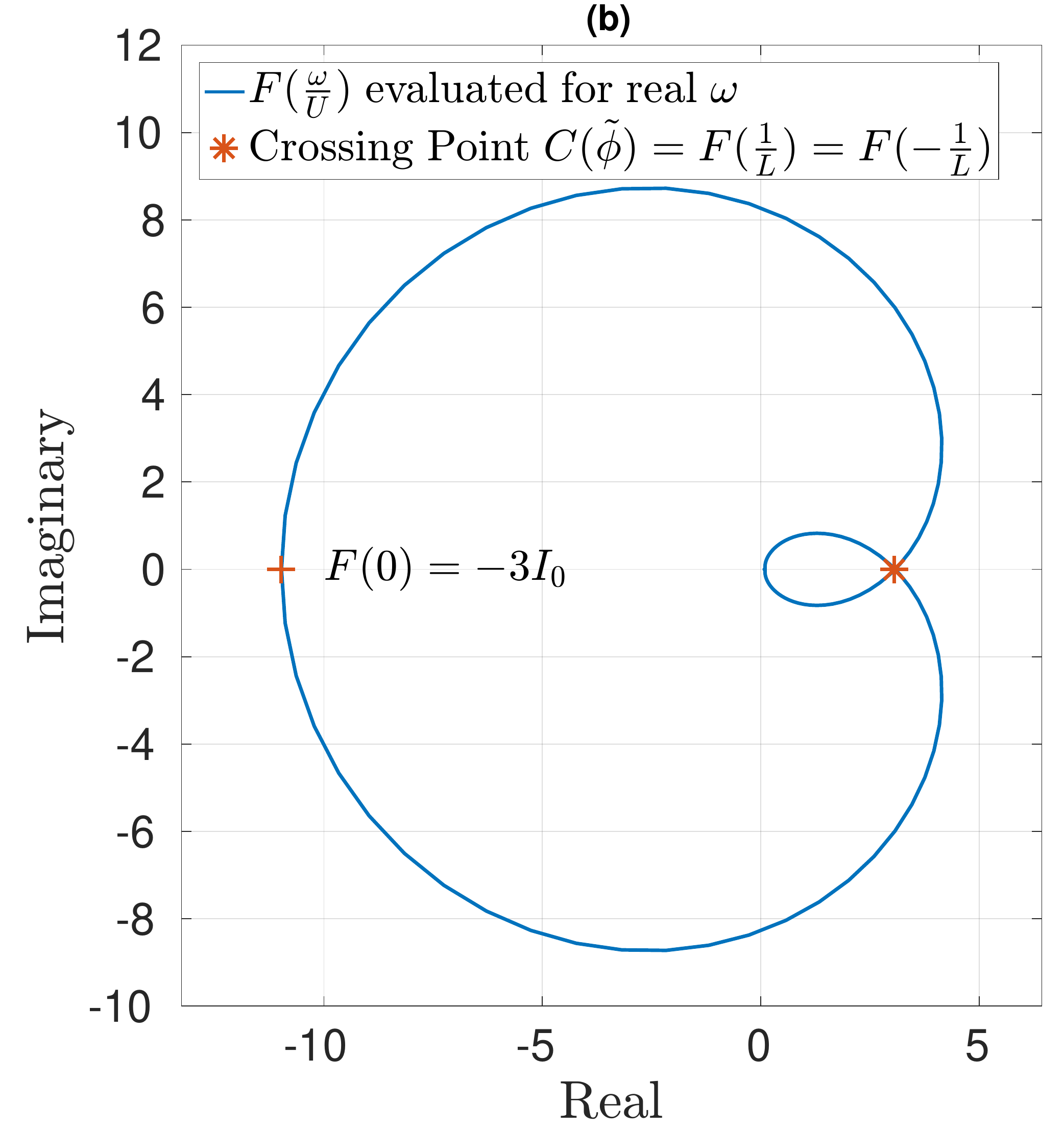}
\caption{(a): $\dot{P}_i/\dot{P}_e$ evaluated on the real axis for $\phi=0.23\sech^4(x/4)\,,m_i/m_e=1836$ and three different hole speeds. $\dot{P}_i/\dot{P}_e(\omega)+1=0$ has two unstable zeros when $|U|<U_c=4.6c_s$ here. (b): $F(\omega/U)$ function defined in Eqn. \eqref{eqn:F} evaluated for $\omega$ on the real axis using $\tilde{\phi}(x)=\sech^4(x/4)$. $F$ contour is invariant for different hole velocity $U$.}
\label{instability_nyquist}
\end{figure}

The equation can be rewritten as $\dot{P}_i/\dot{P}_e+1=0$, where $\dot{P}_i/\dot{P}_e$ is given by a long integral expression
\begin{eqnarray} \label{eqn:full_eqn}
\begin{aligned}
\displaystyle \frac{\dot{P}_i}{\dot{P}_e}(\omega,U,\phi)=&-\displaystyle \frac{m_i}{m_e}\bigg [\displaystyle \int_{x_a}^{x_b}\left(-1-2\displaystyle \frac{U}{v_0(x)}-\displaystyle \frac{U^2}{v_0^2(x)}\right)\exp\left(i\omega\displaystyle \int_{x}^{x_b}\displaystyle \frac{dx_3}{v_0(x_3)}\right)\,dx\\
&\hspace{1cm}-\displaystyle \int_{x_a}^{x_b} \displaystyle \frac{U^2}{v_0^3(x)} \displaystyle \int_{x_a}^{x}\displaystyle \frac{\phi'(x_1)}{v_0(x_1)}\exp\left(i\omega\displaystyle \int_{x_1}^{x_b}\displaystyle \frac{dx_3}{v_0(x_3)}\right)\,dx_1\,dx\\ 
&\hspace{1cm}-\displaystyle \int_{x_a}^{x_b}\displaystyle \int_{x_a}^{x}\displaystyle \frac{U}{v_0^3(x_1)}\displaystyle \int_{x_a}^{x_1}\exp\left(i\omega\displaystyle \int_{x_2}^{x}\displaystyle \frac{dx_3}{v_0(x_3)}\right)\phi''(x_2)\,dx_2\,dx_1\,dx\bigg ]\\
&\bigg/\left(\displaystyle \int_{x_a}^{x_b}h(\sqrt{\phi(x)})+1-\frac{\abs{U}}{\sqrt{U^2-2\phi(x)}}\,dx\right).
\end{aligned}
\end{eqnarray}
$\dot{P}_i/\dot{P}_e$ can be expanded to give a much simpler form in the limit where the ion kinetic energy in the hole frame is much greater than their electrostatic potential energy $U^2\gg 2\psi$ with $\psi$ being the maximum of $\phi$. This approximation is very well satisfied at the onset of instability observed in our simulation. The leading term of the expanded form is
\begin{equation} \label{eqn:expansion}
\frac{\dot{P}_i}{\dot{P}_e}(\omega,U,\phi)
\,\simeq\,-\frac{m_i}{m_e}\frac{\psi^2}{U^4}\frac{4i\displaystyle \frac{\omega}{U}I(\displaystyle \frac{\omega}{U})+i\displaystyle \frac{\omega^2}{U^2}I'(\displaystyle \frac{\omega}{U})-3I_0}{\displaystyle \int_{x_a}^{x_b}h(\sqrt{\phi(x)})-\displaystyle \frac{\phi(x)}{U^2}\,dx},
\end{equation}
where 
\begin{eqnarray}
&&I_0=\int_{x_a}^{x_b}\tilde{\phi}(x)^2\,dx, \\
&&I(\frac{\omega}{U})=\int_{x_a}^{x_b}\int_{x_a}^{y}\tilde{\phi}(x)\tilde{\phi}(y)\exp(i\frac{\omega(x-y)}{U})\,dx\,dy,
\end{eqnarray}
with $\tilde{\phi}(x)=\phi(x)/\psi$ being the normalized potential profile function. This leading term is second order in the small expansion parameter $2\psi/U^2$. The details of this expansion are given in Appendix B. Both the full expression and the leading order expansion of $\dot{P}_i/\dot{P}_e(\omega)$ can be evaluated numerically for real frequencies $\omega$ using for example the widely cited Schamel type of EH potential $\phi(x)=\psi\sech^4(x/4)$. In the evaluation, we use the sign convention that $x_a$ is $-\infty$ and $x_b$ is $+\infty$ in the hole frame. The resulting contours are plotted in the panel (a) of Figure \ref{instability_nyquist}. The number of encirclements of the point $-1$ in the complex plane by the $\dot{P}_i/\dot{P}_e(\omega)$ contour gives the number of unstable $\omega$ solutions to the eigenmode Eqn. \eqref{eqn:eigenmode}. There is a critical speed $U_c$ for $|U|$ below which the system is unstable. The leading order term is within a few percent of the full expression at the onset of instability. From now on, we will work with the leading order term instead of the full expression for the purpose of studying this instability. This approximation makes the mathematics much more tractable. Our analysis is general and can be applied to any type of equilibrium EH potential, including but not limited to Schamel type of EHs.

The $\dot{P}_i/\dot{P}_e(\omega)$ contour is essential to the study of this instability. We are going to take advantage of its scaling property to solve for the critical speed $U_c$. To simplify the notations, we introduce two auxiliary functions $F$ and $G$ defined as
\begin{eqnarray}
&&F(\displaystyle\frac{\omega}{U})\,=\,4i\displaystyle\frac{\omega}{U}I( \displaystyle\frac{\omega}{U})+i \displaystyle\frac{\omega^2}{U^2}I'(\displaystyle\frac{\omega}{U})-3I_0,\label{eqn:F}\\
&&G(U)\,=\,\displaystyle \frac{m_e}{m_i}\displaystyle\frac{1}{\psi^2}\bigg[U^4\displaystyle\int_{x_a}^{x_b} h(\sqrt{\phi(x)})\,dx-U^2\displaystyle\int_{x_a}^{x_b}\phi(x)\,dx\bigg]. \label{eqn:G}
\end{eqnarray}
We have therefore 
\begin{equation}
\displaystyle\frac{\dot{P}_i}{\dot{P}_e}(\omega,U)=-\displaystyle\frac{F(\displaystyle\frac{\omega}{U})}{G(U)}.
\end{equation}
We look for the critical speed $U_c$ for a given equilibrium hole potential $\phi$ such that the $\dot{P}_i/\dot{P}_e(\omega)$ contour crosses the point $-1$ in the complex plane. $F$ depends on $U$ through $\omega/U$, it gives the same contour for different $U$ values when $\omega$ is evaluated on the real axis, although at different $\omega$ values. While $F$ is a complex-valued function, function $G$ only takes real values. $U$ scales the size of the $\dot{P}_i/\dot{P}_e(\omega)$ contour through $G$. This property is demonstrated in Figure \ref{instability_nyquist}. The identical $F(\omega/U)$ contour for different values of $U$ using a Schamel type of EH potential is shown in the panel (b) of Figure \ref{instability_nyquist}. We denote its intersection with the positive real axis by $C(\tilde{\phi})$. The existence of this intersection $C(\tilde{\phi})$ is guaranteed for a general class of admissible hole potential $\phi(x)$, which we will show later in this paper. The critical speed $U_c$, below which the system is unstable, satisfies an equation
\begin{equation}
-\frac{C(\tilde{\phi})}{G(-U_c)}\,=\,-1.
\end{equation}
The above equation gives a quadratic equation in $U_c^2$
\begin{equation} \label{eqn:critical_velocity}
U_c^4\int_{x_a}^{x_b} h(\sqrt{\phi(x)})\,dx-U_c^2\int_{x_a}^{x_b}\phi(x)\,dx-\psi^2\frac{m_i}{m_e}C(\tilde{\phi})=0.
\end{equation}
The unique real and positive solution of $U_c^2$ is
\begin{equation}\label{eqn:U_c}
U_c^2\,=\,\displaystyle\frac{\displaystyle\int_{x_a}^{x_b}\phi(x)\,dx+\sqrt{\left(\displaystyle\int_{x_a}^{x_b}\phi(x)\,dx\right)^2+4\psi^2\displaystyle\frac{m_i}{m_e}C(\tilde{\phi})\displaystyle\int_{x_a}^{x_b}h(\sqrt{\phi(x)})\,dx}}{2\displaystyle\int_{x_a}^{x_b}h(\sqrt{\phi(x)})\,dx}.
\end{equation}

Now we calculate the oscillation frequency of the unstable eigenmode. At marginal instability, the $\dot{P}_i/\dot{P}_e(\omega)$ contour crosses $-1$. We need to find the frequency for which this crossing happens. The imaginary part of $\dot{P}_i/\dot{P}_e$ is
\begin{equation}
\mathrm{Im}\left(\frac{\dot{P}_i}{\dot{P}_e}(\omega)\right)\,=\,-\frac{\mathrm{Im}\left(F(\displaystyle\frac{\omega}{U})\right)}{G(U)},
\end{equation}
where
\begin{equation}
\mathrm{Im}\left(F(\displaystyle\frac{\omega}{U})\right)\,=\,4\frac{\omega}{U}\mathrm{Re}\left(I(\frac{\omega}{U})\right)+\frac{\omega^2}{U^2}\mathrm{Re}\left(I'(\frac{\omega}{U})\right).
\end{equation}
To calculate the imaginary part of $\dot{P}_i/\dot{P}_e$, we need to evaluate $\mathrm{Re}(I(\frac{\omega}{U}))$. It can be shown by taking $x_a$ and $x_b$ to infinity that
\begin{equation} \label{eqn:Fourier}
\begin{aligned}
\mathrm{Re}\left(I(\frac{\omega}{U})\right)\,&=\,\frac{1}{2}\left(\int_{-\infty}^{\infty}\tilde{\phi}(x)\exp\left(i\frac{\omega}{U}x\right)\,dx\right)\left(\int_{-\infty}^{\infty}\tilde{\phi}(y)\exp\left(-i\frac{\omega}{U}y\right)\,dy\right)\\
&=\,\frac{1}{2}\tilde{\Phi}(\frac{\omega}{U})^2,
\end{aligned}
\end{equation}
where $\tilde{\Phi}$ is the modulus of the Fourier transform of $\tilde{\phi}$. Thus $\mathrm{Im}(\dot{P}_i/\dot{P}_e)(\omega)=0$ gives
 \begin{equation} \label{eqn:Imag_part}
\begin{aligned}
\mathrm{Im}\left(F(\displaystyle\frac{\omega}{U})\right)\,\equiv\,\frac{\omega}{U}\tilde{\Phi}(\frac{\omega}{U})(2\tilde{\Phi}(\frac{\omega}{U})+\frac{\omega}{U}\tilde{\Phi}'(\frac{\omega}{U}))\,=\,0
\end{aligned}
\end{equation}
Eqn. \eqref{eqn:Imag_part} admits three real solutions for $\omega$, one is the trivial $\omega=0$, the other two solutions are given by the equation
\begin{equation} \label{eqn:p}
-\frac{\tilde{\Phi}'(\displaystyle\frac{\omega}{U})}{2\tilde{\Phi}(\displaystyle\frac{\omega}{U})}\,=\,\frac{1}{\omega/U}.
\end{equation}
Since $\tilde{\phi}$ is real, we have $\tilde{\Phi}$ is an even function and Eqn. \eqref{eqn:p} gives two solutions that have the opposite sign. We define the positive $\omega$ solution of Eqn. \eqref{eqn:p} as $\omega_0$ and a length scale $L$
\begin{equation}
L\,\equiv\,\frac{\abs{U}}{\omega_0}.\label{eqn:L}
\end{equation}
The $\dot{P}_i/\dot{P}_e$ contour crosses the real axis at frequency $\omega_0$ for a given EH potential $\phi$ and $U$. $L$ is the characteristic length of the EH and it is entirely determined by the EH potential \emph{shape} $\tilde{\phi}$. At the onset of instability, we have $\dot{P}_i/\dot{P}_e(\omega_0(U_c),-U_c)=-1$, $\omega_0$ evaluated for the critical speed $U_c$ is therefore the angular frequency of the initially growing unstable eigenmode. We define this frequency as 
\begin{equation} \label{eqn:omega_c}
\omega_c\,\equiv\,\omega_0(U_c)\,\equiv\,\frac{U_c}{L}.
\end{equation}
It is the critical ion transit frequency through the hole potential. The frequency of the growing oscillation corresponds to a physical frequency of the system.

The existence of this critical frequency $\omega_c$ and the crossing point $C(\tilde{\phi})$ are guaranteed by the continuous differentiablility of the EH potential $\phi(x)$. A physical EH potential $\phi(x)$ should possess a second derivative as it satisfies Poisson's equation $\phi''(x)+\rho(x)/\epsilon_0=0$, and a physical $\rho(x)$ should have bounded variation. This smoothness requirement constrains the asymptotic behavior of its Fourier transform. Function $\tilde{\Phi}(p)$ decays at least as fast as $p^{-3}$ when $p\rightarrow \infty$ \citep{Katznelson02}. Thus we have $-\tilde{\Phi}'(p)/2\tilde{\Phi}(p)=-\frac{1}{2}d\ln(\tilde{\Phi}(p))/dp\geq 3/2p>1/p$ as $p\rightarrow \infty$. While as $p\rightarrow 0$, we have $-\tilde{\Phi}'(p)/2\tilde{\Phi}(p)\rightarrow -\tilde{\Phi}'(0)/2\tilde{\Phi}(0) \ll1/p$. Solutions are guaranteed for Eqn. \eqref{eqn:p}. The behavior of the $F$ contour is as follows. As $\omega/U\rightarrow 0^{+}$, we have $\mathrm{Re}(F(\omega/U))\rightarrow -3I_0<0$ and $\mathrm{Im}(F(\omega/U))\rightarrow 0^{+}$. Asymptotic analysis as $\omega/U\rightarrow +\infty$ gives $\mathrm{Re}(F(\omega/U))\rightarrow 0^{+}$ and $\mathrm{Im}(F(\omega/U))\rightarrow 0^{-}$. In other words, with $\omega/U$ increasing from $0$ to infinity, the $F$ contour starts from a point on the negative real axis, goes into the upper half plane, crosses the positive real axis at $1/L$ and returns to zero. The crossing point $C(\tilde{\phi})$ shown in Figure \crefformat{figure}{#2#1{(b)}#3}\cref{instability_nyquist}, which is crucial to this instability, is an universal feature for all physically admissible hole potentials $\phi(x)$ and we have $C(\tilde{\phi})=F(1/L)$. Contours without a crossing can be obtained only from unphysical hole shapes. For example, $\tilde{\phi}=\exp(-\abs{x}/\lambda)$ does not give a crossing point and is therefore stable; but it is unphysical, as the electric field is undefined at $x=0$.
\begin{figure}[hbt]
\centering
\includegraphics[height=0.55\textwidth]{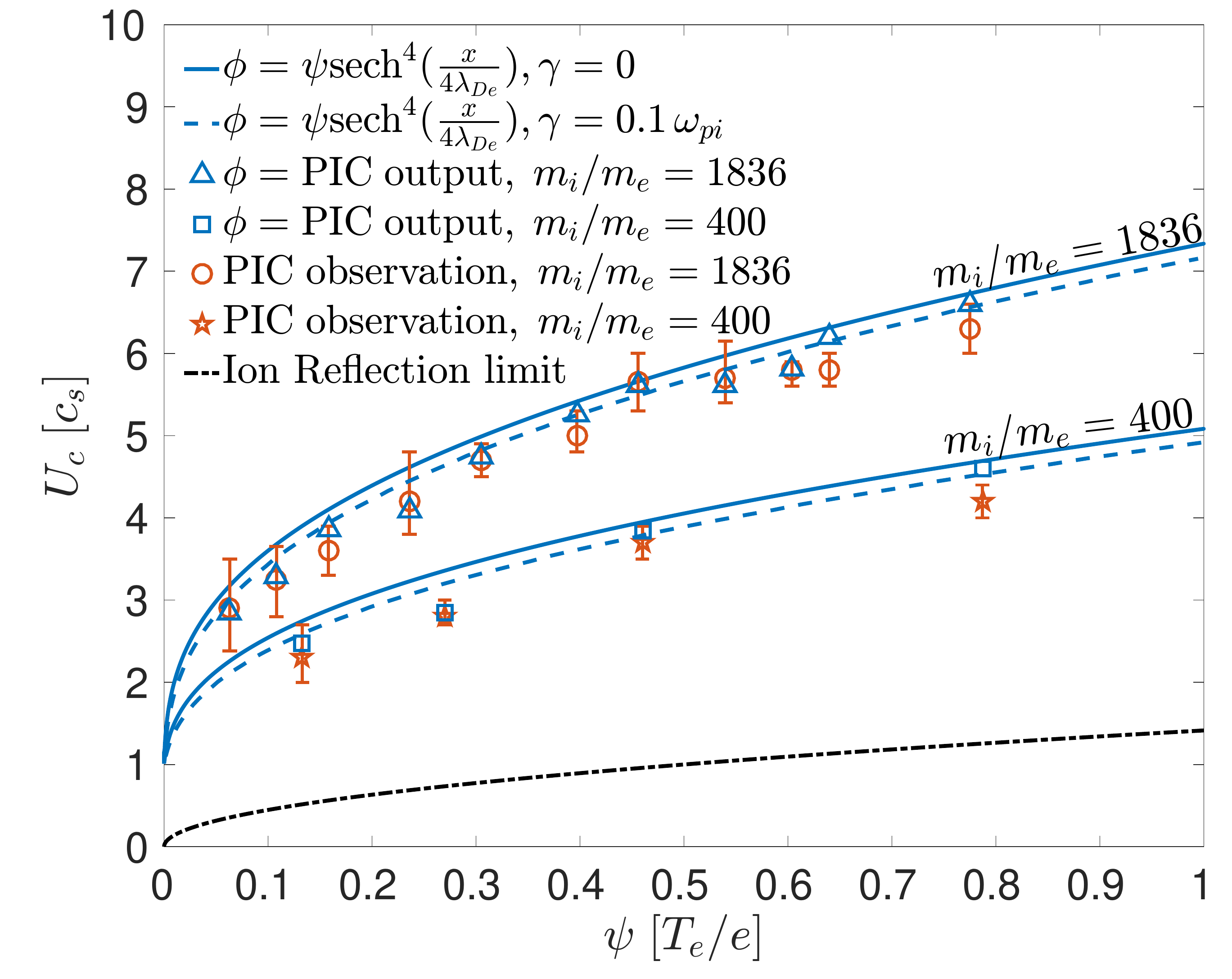}
\caption{The critical values of hole speed in the ion frame below which the instability occurs for different sized EHs and two different mass ratios. The theoretical stability boundaries ($\gamma=0$) and the $\gamma=0.1$ growth rate boundaries for Schamel type of EHs $\phi(x)=\psi\sech^4(x/4)$ are plotted as reference lines. The observational data point and the numerical calculation of the same $\psi$ correspond to the same run. The ion reflection limit is much lower than the instability threshold, hence our approximation $U^2\gg 2\psi$ is well satisfied. All the PIC runs have $T_e/T_i=20$.}
\label{instability_boundary}
\end{figure}
\begin{figure}[hbt]
\centering
\includegraphics[height=0.55\textwidth]{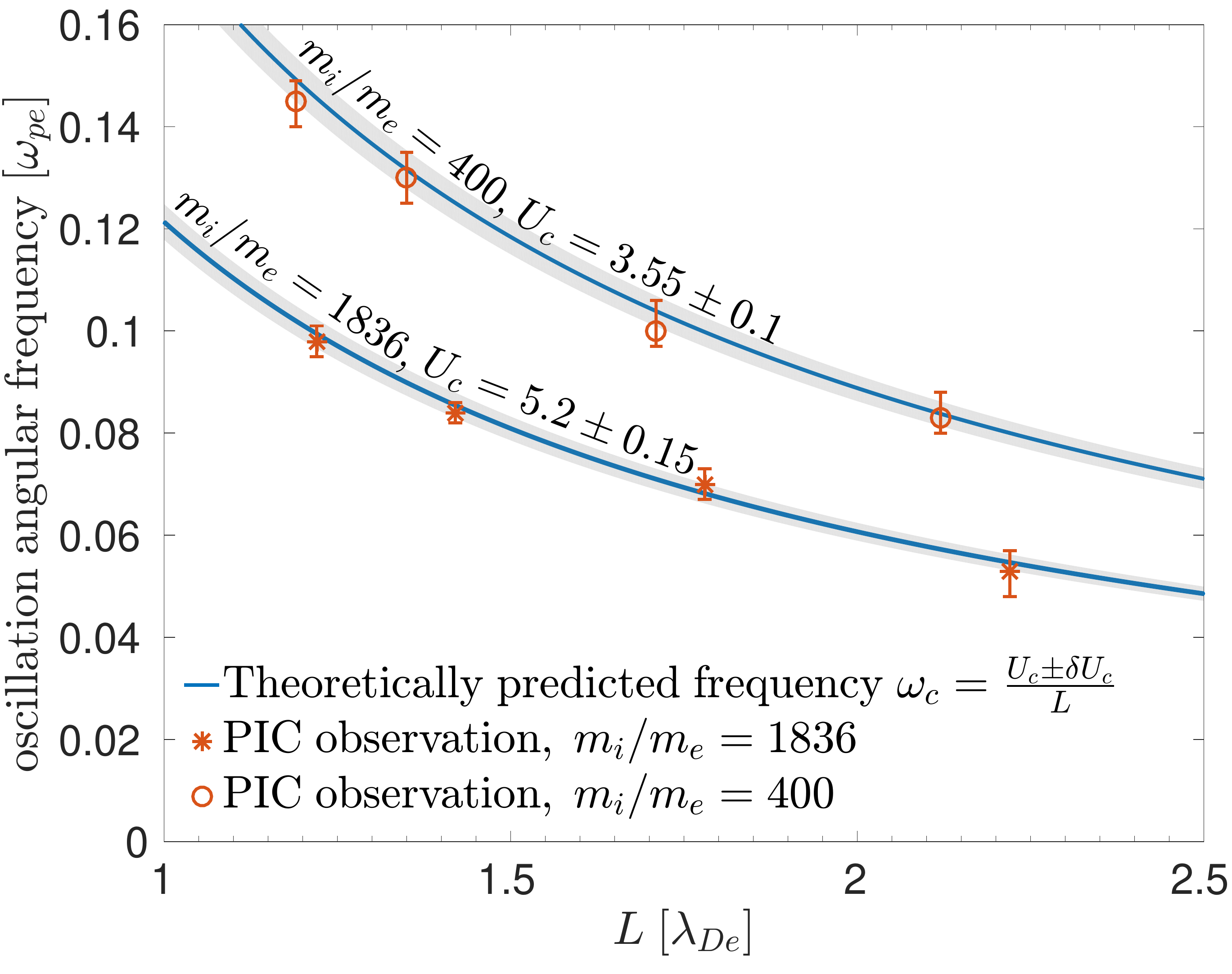}
\caption{The oscillations seen in our simulation are Fourier analyzed to extract its main frequency for the first few periods of unstable oscillations. The uncertainty in the theoretically predicted frequency due to the uncertainty of $U_c$ used in Eqn. \eqref{eqn:omega_c} is shown by the gray uncertainty bands. Notice that the unstable oscillation frequency is in general a few times the ion plasma frequency.}
\label{instability_frequency}
\end{figure}

Having obtained the analytic solution for the critical speed and the unstable oscillation frequency, we now compare these results with our PIC observations. The hole pushing technique enables us to explore continuously the EH velocity in the ion frame. We performed a series of runs with different initialization to create EHs of different sizes. Then we determined the critical speed $U_c$ by inspecting the onset of unstable velocity oscillations as the hole speed is decreased. It is compared with the threshold speeds obtained by solving the Nyquist stability problem numerically using the $\phi(x)$ right before the instability onset from the same run. The electrostatic potential output $\phi(x)$ from our PIC simulation is used to construct the $F$ contour numerically from Eqn. \eqref{eqn:F} and find its crossing point $C(\tilde{\phi})$. We use the formula for $U_c$ in Eqn. \eqref{eqn:U_c} to calculate its predicted value. This method takes into account the exact potential shape of the EH in our PIC simulation which is different from one run to another \footnote{Our hole-tracking PIC simulation produces relatively low-noise and highly resolved EH potential. We applied some post processing to the PIC potential output to make the numerical calculation more accurate. In our analysis, $\phi(x)$ is considered to fall to zero far away from the hole center. However, there is always some non-zero intrinsic statistical noise in the PIC simulation. In post processing, we find the positions where the electric field first becomes zero outside the hole center and consider them to be the limits of the hole spatial extent. The values of $\phi(x)$ beyond these limits are forced to decay to zero by multiplying a Debye decaying exponential to them. We use this slightly smoothed $\phi(x)$ in our numerical calculation of $U_c$.}. The results are presented in Figure \ref{instability_boundary}. The solid lines are obtained using Eqn. \eqref{eqn:U_c} assuming a Schamel type of EH potential. This solution's asymptotic behavior comes from the special function $h(\chi)$ \cite{Hutchinson2016a}: $h(\chi)\rightarrow \chi^2-\frac{8}{3\sqrt{\pi}}\chi^3$ as $\chi\rightarrow 0$ and $h(\chi)\rightarrow 1-\frac{2}{\sqrt{\pi}\chi}$ as $\chi\sim 1$. For shallow holes $\psi\ll 1$, we have $U_c\simeq1+\mathcal{O}(\psi)$ and for deep holes $\psi\sim1$, $U_c\sim(m_i/m_e)^{1/4}\psi^{1/2}$. The slight deviation of our data points from the solid curves represents the deviation of the hole potential in our PIC simulation from the Schamel type. The full calculation using the exact hole potential yields a good agreement with the observation. We have runs with two different mass ratios and our results show the $(m_i/m_e)^{1/4}$ scaling of $U_c$ with the mass ratio as predicted by the theory. The linear growth rate $\gamma=\mathrm{Im}(\omega)$ of the instability when $\abs{U}<U_c$ can be evaluated by solving the equation $\dot{P}_i/\dot{P}_e(\omega)+1=0$ numerically for given $\phi(x)$, $U$ and the mass ratio. In Figure \ref{instability_boundary}, we show in dashed lines the EH velocity calculated as a function of $\psi$ for $\gamma=0.1$ and Schamel type of EH potential as a useful reference line. We shall give a detailed analysis of the growth rate in \ref{subsec:gamma}.

In Figure \ref{instability_frequency}, we show the frequencies of the unstable velocity oscillations seen in our simulation. The EH position and hence velocity $U$ is obtained from the hole-tracking module for each PIC time step of $0.3/\omega_{pe}$. A discrete Fourier transform of $U$ during the first few unstable periods gives a sharp peak centered at the oscillation frequency. The error bar is given by the frequency range above half peak power. They are plotted against the characteristic hole width $L$ defined in Eqn. \eqref{eqn:L} calculated using the potential $\phi(x)$ right before the oscillation onset from our PIC simulations. $\phi(x)$ gives numerically the $F$ contour and it crosses the positive real axis at $F(1/L)$. EH initialization is adjusted in our PIC code to give EHs of different width $L$ and a narrow range of potential height $\psi\sim0.45$, hence $U_c$. The oscillation frequency predicted by our theory is inversely proportional to the hole width $L$: $\omega_c=U_c/L$. Good quantitative agreement is achieved between the observations and our theory. Our analysis captures the correct scaling with the mass ratio, which highlights the importance of  ion dynamics for this instability.

The instability threshold $U_c$ is scale invariant. If we apply a change of scale $x\rightarrow\lambda x$, the Eqn. \eqref{eqn:critical_velocity} giving the critical speed $U_c$ is invariant under this change of scale as each term is multiplied by the same factor $1/\lambda$. This property is obvious for the first two terms in Eqn. \eqref{eqn:critical_velocity}. For the third term, it can be shown that 
\begin{equation}
F_{\lambda}(\frac{\omega}{U})\,=\,\frac{1}{\lambda}F(\frac{\omega}{U\lambda}).
\end{equation}
Hence the crossing point satisfies the same scaling relation $C_{\lambda}(\tilde{\phi})=(1/\lambda)C(\tilde{\phi})$ and $U_c$ remains invariant. However, the oscillation frequency scales linearly with $\lambda$: $\omega_{c,\lambda}=\lambda\omega_{c}$. For example, two different EH potentials $\phi(x)=\psi\sech^4(x)$ and $\phi(x)=\psi\sech^4(x/4)$ have the same threshold $U_c$, while the unstable oscillation frequency for the first potential profile is four times as high. This argument explains why the runs in Figure \ref{instability_frequency} have a similar $U_c$ but different oscillation frequencies.
\subsection{Counter-streaming ions}
We have shown an example of the instability observed in a plasma with counter propagating ions in Figure \ref{instability_CounterStreaming}. If the EH potential $\phi(x)$ is symmetric, then the counter-streaming situation with an EH at rest $U=0$ and two ion streams traveling at $\pm v_i$ is equivalent to having one single ion stream at rest and the EH traveling at $U=v_i$ for the described instability mechanism. The sign convention is our analysis is such that the ions enter from $-\infty$ and exit at $+\infty$. The change of hole velocity from $U$ to $-U$ in the ion frame results in flipping the sign convention thus $x_a$ and $x_b$. When the potential $\phi(x)$ is symmetric so that $\phi'(-x)=-\phi'(x)$ and $\phi''(-x)=\phi''(x)$, Eqns. \eqref{eqn:io} \eqref{eqn:contained} show that the resulting $\dot{P}_i$ is exactly the opposite as $\dot{U}$ has an opposite sign under the two opposite sign conventions. Therefore, the contribution to the total $\dot{P}_i$ from the two ion streams, evaluated with the same sign convention, should be exactly equal and add up. More concretely, an ion particle arriving from the left sees the same potential as an ion particle arriving from the right. However, the same EH acceleration $\dot{U}$ works in an opposite way for them. This argument explains why the instability threshold observed in the counter-streaming ion plasma is identical to the threshold value for the single ion stream case. The two situations are equivalent in terms of linear stability. Once the instability has fully grown, the nonlinear stage of the instability can be different for the two cases.
\subsection{Linear growth rate}\label{subsec:gamma}
The linear growth rate of the instability is obtained by solving the eigenmode equation $\dot{P}_i/\dot{P}_e(\omega,U)+1=0$. The growth rate $\gamma$ is the imaginary part of the solution $\omega$: $\gamma=\mathrm{Im}(\omega)$. Although analytic solution for arbitrary $U$ is too difficult, we can obtain $\gamma$ by an expansion near marginal instability. Recall that at marginal instability, we have
\begin{equation}
\frac{\dot{P}_i}{\dot{P}_e}(\omega_c=\frac{U_c}{L},-U_c)\,=\,-1.
\end{equation}
If the hole velocity is $U=-U_c+\Delta U$ such that $|\Delta U/U_c|\ll 1$. We need to find $\omega=\omega_c+\Delta \omega$ with $|\Delta \omega/\omega_c|\ll 1$ that satisfies the eigenmode equation $\dot{P}_i/\dot{P}_e(\omega,U)=-1$. A linear expansion gives
\begin{equation} \label{eqn:linear_eigenmode}
\Delta \omega \frac{\partial (\dot{P}_i/\dot{P}_e)}{\partial \omega}\bigg|_{\omega_c,-U_c}+\Delta U \frac{\partial (\dot{P}_i/\dot{P}_e)}{\partial U}\bigg|_{\omega_c,-U_c}\,=\,0.
\end{equation}
Substituting $\dot{P}_i/\dot{P}_e(\omega,U)=-F(\omega/U)/G(U)$, the two partial derivatives in Eqn. \eqref{eqn:linear_eigenmode} can be evaluated with functions $F$ and $G$
\begin{eqnarray}
&& \frac{\partial (\dot{P}_i/\dot{P}_e)}{\partial \omega}\bigg|_{\omega_c,-U_c}\,=\,\frac{F'(-\omega_c/U_c)}{U_cG(-U_c)},\\
&&\frac{\partial (\dot{P}_i/\dot{P}_e)}{\partial U_c}\bigg|_{\omega_c,-U_c}\,=\,\frac{F'(-\omega_c/U_c)G(-U_c)(\omega_c/U_c^2)+F(-\omega_c/U_c)G'(-U_c)}{G(-U_c)^2}.
\end{eqnarray}
Hence
\begin{equation}
\begin{aligned}
\Delta \omega\,=&\,\Delta U\bigg\{-\frac{\omega_c}{U_c}-U_c\frac{F(-\omega_c/U_c)}{G(-U_c)}\frac{G'(-U_c)}{F'(-\omega_c/U_c)}\bigg\}\\
=&\,\Delta U\bigg\{-\frac{\omega_c}{U_c}-U_c\frac{G'(-U_c)}{F'(-\omega_c/U_c)}\bigg\},
\end{aligned}
\end{equation}
where we used $F(-\omega_c/U_c)/G(-U_c)=1$. $G$ is an even polynomial function defined in Eqn. \eqref{eqn:G} and $-U_cG'(-U_c)\equiv U_cG'(U_c)$ can be evaluated as
\begin{equation}
\begin{aligned}
U_cG'(U_c)\,=\,4F(\frac{\omega_c}{U_c})+2U_c^2\frac{m_e}{m_i}\displaystyle\frac{1}{\psi^2}\displaystyle\int_{x_a}^{x_b}\phi(x)\,dx.
\end{aligned}
\end{equation}
Hence the final expression for $\Delta \omega$ is
\begin{equation}\label{eqn:delta_omega}
\Delta \omega\,=\,\Delta U\bigg\{-\frac{\omega_c}{U_c}+\frac{4F(\omega_c/U_c)}{F'(-\omega_c/U_c)}+2U_c^2\frac{m_e}{m_i}\frac{1}{F'(-\omega_c/U_c)}\displaystyle\frac{1}{\psi^2}\displaystyle\int_{x_a}^{x_b}\phi(x)\,dx\bigg\}.
\end{equation}
The growth rate $\gamma$ is the imaginary part of $\Delta \omega$. The real part of $\Delta \omega$ gives a small correction to the oscillation frequency $\omega_c$ when $U$ is different from $U_c$. We also have $\omega_c/U_c=1/L$, a constant only depending on the hole shape, and $F(\omega_c/U_c)=F(1/L)=C(\tilde{\phi})$. While $F(\omega_c/U_c)$ is a real number, the derivative $F'(-\omega_c/U_c)$ is complex and $\gamma$ is given by
\begin{equation}
\gamma\,=\,\Delta U\, \mathrm{Im}\bigg\{\frac{4F(1/L)}{F'(-1/L)}+2U_c^2\frac{m_e}{m_i}\frac{1}{F'(-1/L)}\displaystyle\frac{1}{\psi^2}\displaystyle\int_{x_a}^{x_b}\phi(x)\,dx\bigg\}.
\end{equation}
The first term is only a function of the hole shape $\tilde{\phi}$ while the second term depends on hole size $\psi$ and the mass ratio $m_i/m_e$. However, this second term is not important except for extremely shallow EHs such that $\psi \ll 1$. For example, for a Schamel type of EH of size $\psi=0.1$ and $m_i/m_e=1836$, the magnitude of the second term is about $4\%$ of the first one. It is thus a good approximation that for not too shallow EHs we have
\begin{equation} \label{eqn:gamma}
\gamma\,\simeq\,\Delta U\, \mathrm{Im}\bigg\{\frac{4F(1/L)}{F'(-1/L)}\bigg\}\,=\,\Delta U\, \mathrm{Im}\bigg\{\frac{4F(1/L)}{F'(1/L)}\bigg\}\,=\,-\Delta |U|\, \mathrm{Im}\bigg\{\frac{4F(1/L)}{F'(1/L)}\bigg\}.
\end{equation}
We define $\Delta|U|=|U|-U_c$ and the imaginary part of $F$  is odd so its derivative is even: $\mathrm{Im}(F'(-1/L))=\mathrm{Im}(F'(1/L))$. This growth rate scales linearly with $\Delta \abs{U}$. If the hole potential shape is of Schamel type, a numerical evaluation of the constants gives $\gamma\simeq-\Delta \abs{U}/1.74$ for $\abs{U}$ evaluated in $c_s$ and $\gamma$ evaluated in $\omega_{pi}$. The instability grows fast once $\abs{U}$ is slower than $U_c$.

\begin{figure}[hbt]
\centering
\includegraphics[width=0.7\textwidth]
{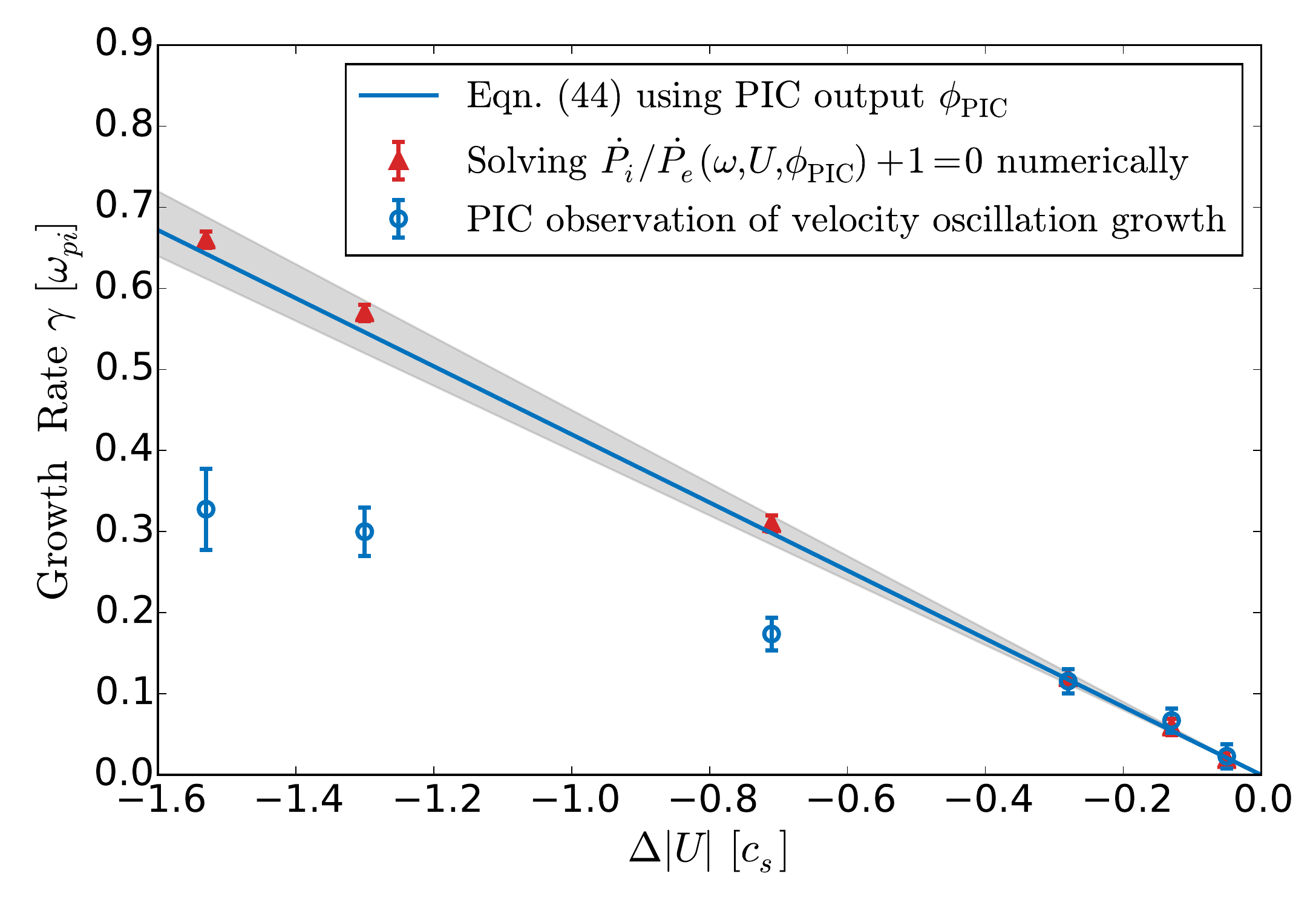}
\caption{Instability growth rate $\gamma$ as a function of $\Delta U$. The line represents Eqn. \eqref{eqn:gamma}  for fixed hole shape. Its uncertainty bands represent the small variation of shape from one run to another, giving uncertainty in the comparison. The triangles are obtained from solving numerically the full eigenmode equation $\dot{P}_i/\dot{P}_e+1=0$ using the PIC potential output. Circles are the growth rate observed in PIC runs.}
\label{linear_growth}
\end{figure}

In the PIC simulations, we measured the growth rate of unstable velocity oscillations by fitting an exponential growth model to it. We used the unstable runs with different counter-streaming ion velocity and $\psi\sim0.8$, in which $\Delta \abs{U}$ can be precisely measured. The growth rates and the error bars are obtained from the regression. They are compared with the expanded linear solution in Eqn. \eqref{eqn:gamma} and numerical solutions of the full eigenmode equation. The results are shown in Figure \ref{linear_growth}. Our analytic theory agrees with the observed instability growth rate for the weakly unstable cases up to $\gamma\sim0.1$ and the linear expansion gives a very good approximation to the solution of the full eigenmode equation. The growth rates for strongly unstable cases are less than the predicted values. We attribute this discrepancy to our inability to observe the growth rate in a truly linear stage when $\gamma$ is large. We are going to discuss these nonlinear effects in Section \ref{sec:3}.

\subsection{Finite ion temperature}
So far, we have treated the ions as a cold beam. In reality, they have a thermal velocity spread. For an ion of velocity $v$, the EH has a velocity $U-v$ in its frame. Let's consider the ion thermal speed to be small compared to $\abs{U}$ such that $v_\mathrm{th,i}\ll \abs{U}$ and there are no reflected ions by the hole potential. As $U_c$ is several times $c_s$, this assumption holds approximately even when $T_i\geq T_e$. We can integrate the contributions from ions of different velocities to get the total $\dot{P}_i$
\begin{equation}\label{eqn:P_i_thermal}
\dot{P}_i(\omega,U,\phi,v_\mathrm{th,i})\,=\,m_i\dot{U}\int_{-\infty}^{\infty}f_{\infty,i}(v)\frac{\psi^2}{(U-v)^4}F(\frac{\omega}{U-v})\,dv.
\end{equation}
We apply a Taylor series expansion to Eqn. \eqref{eqn:P_i_thermal} assuming $\abs{v}\ll \abs{U}$. Consider $f_{\infty,i}(v)$ to be a Maxwellian and only the even order moments of $v$ survive after the integration over velocity. This expansion gives
\begin{equation}
\begin{aligned}
\dot{P}_i(\omega,U,\phi,v_\mathrm{th,i})\,&=\,n_\infty m_i\dot{U}\frac{\psi^2}{U^4}\bigg\{F(\frac{\omega}{U})+F_2(\frac{\omega}{U})(\frac{v_\mathrm{th,i}}{U})^2+\mathcal{O}\left((\frac{v_\mathrm{th,i}}{U})^4\right)\bigg\}\\
&=\,n_\infty m_i\dot{U}\frac{\psi^2}{U^4}\bigg\{F_\mathrm{th,i}(\frac{\omega}{U})+\mathcal{O}\left((\frac{v_\mathrm{th,i}}{U})^4\right)\bigg\},
\end{aligned}
\end{equation}
where
\begin{equation}
F_2(\frac{\omega}{U})\,=\,10F(\frac{\omega}{U})+5F'(\frac{\omega}{U})(\frac{\omega}{U})+\frac{1}{2}F''(\frac{\omega}{U})(\frac{\omega}{U})^2.
\end{equation}
We have right now $\dot{P}_i/\dot{P}_e(\omega,U,\phi,v_\mathrm{th,i})=-F_\mathrm{th,i}(\omega/U)/G(U)$ to the leading order in $|v_\mathrm{th,i}/U|$. It suffices to substitute $F$ with $F_\mathrm{th,i}$ in our previous analysis and everything follows as before.

\begin{figure}[hbt]
\centering
\includegraphics[width=0.5\textwidth]
{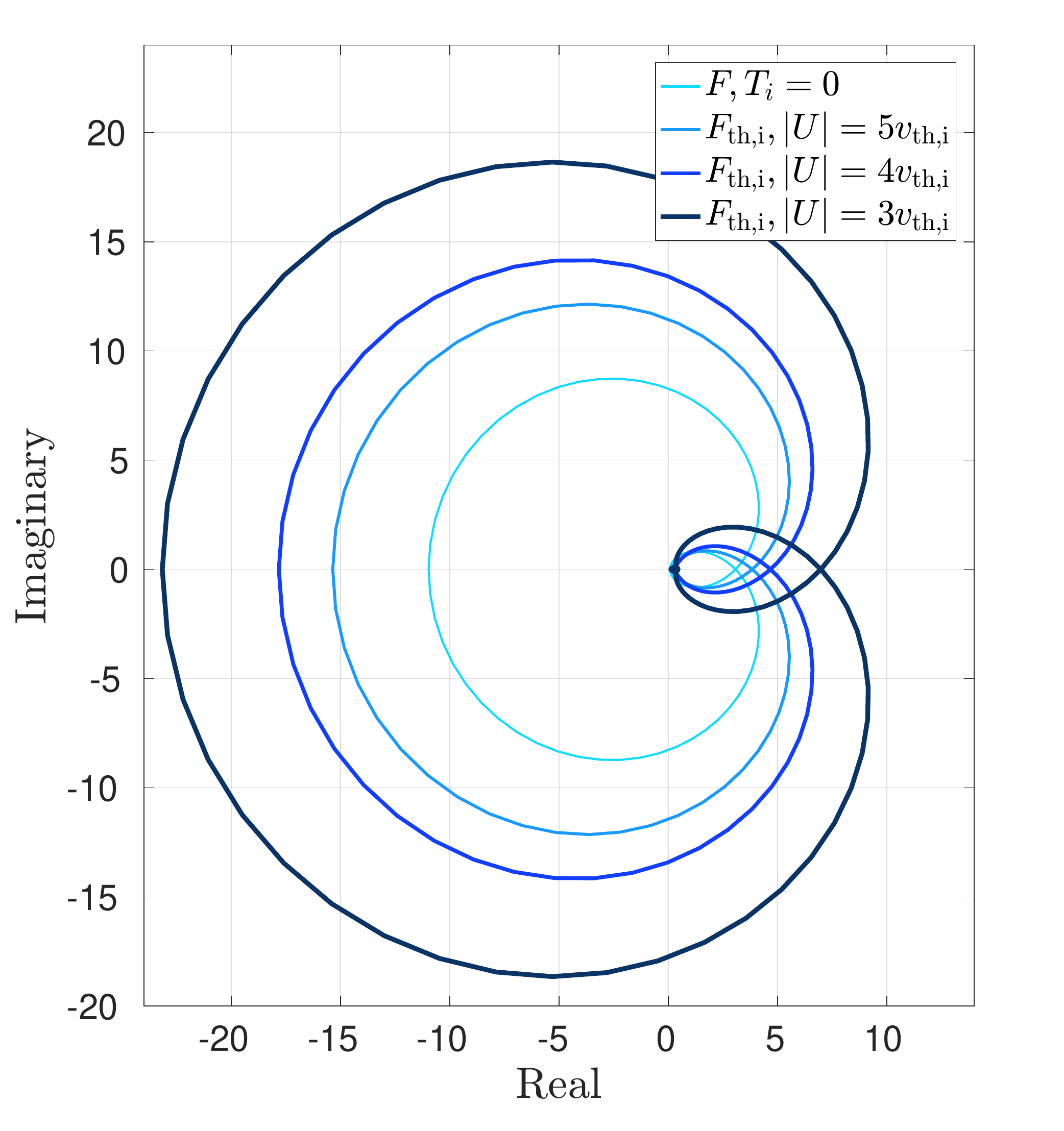}
\caption{Finite ion temperature effect on the $F$ contour for a Schamel type of EH. The contour shape is approximately preserved while its size grows with a larger $T_i$.}
\label{ion_temperature}
\end{figure}

The leading order term of the finite ion temperature correction is second order in $|v_\mathrm{th,i}/U|$. The effect of finite ion temperature can be visualized through the $F_\mathrm{th,i}$ contours. In Figure \ref{ion_temperature}, we show the $F_\mathrm{th,i}$ contours evaluated on the real axis for different values of $|v_\mathrm{th,i}/U|$ using Schamel type of EH potential. The most salient effect is that the finite ion temperature moves the crossing point $C(\tilde{\phi})$ outwards, resulting in a higher value for $U_c$. Because of the leading $U^4$ term in $G(U)$, the resulting change in $U_c$ is actually relatively small. In terms of $U_c$, this correction is $\sim5\%$ when $|U|=5v_\mathrm{th,i}$, it grows to $\sim10\%$ for $|U|=4v_\mathrm{th,i}$ and $\sim20\%$ when $|U|=3v_\mathrm{th,i}$. This property holds similarly for other EH potential models such as the Gaussian. The same trend is noticed in our PIC simulation. A higher ion temperature $T_i$ leads to a slightly higher threshold velocity $U_c$.

When $|U|<3v_\mathrm{th,i}$, the ion reflection from the EH potential becomes important and can no longer be neglected in the global momentum balance. The unbalanced scattering of ions tends to accelerate the EH to a higher velocity in the ion frame \citep{Hutchinson2016a}. With hole pushing technique, we were able to explore the situation with the presence of mild ion reflection from the hole potential, the instability is still observed in these cases. After the onset of instability, we stopped hole pushing and observed the hole velocity to oscillate while its mean velocity accelerates due to the ion reflection.  Resonant ion effects such as ion Landau damping or reflection only become important when $|U|\sim v_\mathrm{th,i}$, which requires the ions to be extremely hot. Our analysis holds well for the usual range of $T_e/T_i$ in space plasmas.

\section{Discussion}\label{sec:3}

The deformation of hole potential during oscillation has been neglected in our analysis. It is a next order correction for the ``jetting" effect we've calculated. We now show that the ion momentum change due to the hole potential variation can be ignored in the parameter regime we are interested in. When an ion transits through the EH potential, its momentum changes when there is a temporal variation of the hole potential height. We call this the hole growth ``jetting" effect and a formula is given in our previous paper \cite{Hutchinson2016a}
\begin{equation}
(v_{bf}-v_{as})_{\mathrm{growth}}\simeq-\frac{1}{U}\int_{x_a}^{x_b}\frac{1}{v}\frac{\partial \phi}{\partial t}\,dx,
\end{equation}
where $\partial \phi/\partial t$ represents the temporal variation of the hole potential in its rest frame. It is related to the change in charge density by Poisson's equation. To a first approximation, we assume that the frequency of velocity oscillation is low so that the electron density remains the same in the hole frame. Therefore, the only density variation comes from the ions and we have approximately $\delta \phi/\phi \sim \delta n_i/n_i$. Using $n_i\partial\phi/\partial t\sim\phi\partial n_i/\partial t$, we get
\begin{equation}
\begin{aligned}
(v_{bf}-v_{as})_{\mathrm{growth}}\sim&\,\frac{1}{U}\int_{x_a}^{x_b}\frac{\phi}{vn_i}\frac{\partial n_i}{\partial t}\,dx\\
\sim&\,\frac{1}{U}\int_{x_a}^{x_b}\frac{\phi}{vn_i}\frac{n_i\dot{U}}{v}\,dx\\
\sim&\,\frac{\psi}{U^2}(v_{bf}-v_{as})_\mathrm{accel}.
\end{aligned}
\end{equation}
The $(v_{bf}-v_{as})_\mathrm{accel}$ is the ``jetting" effect due to hole acceleration. It was shown in Eqn. \eqref{eqn:jetting_term}. The ion momentum change due to self-consistent EH potential variation is on the order of a factor $\psi/U^2\ll 1$ smaller than the momentum change due to EH acceleration. Thus we can ignore it in the momentum balance.

\begin{figure}[hbt]
\centering
\includegraphics[width=0.6\textwidth]
{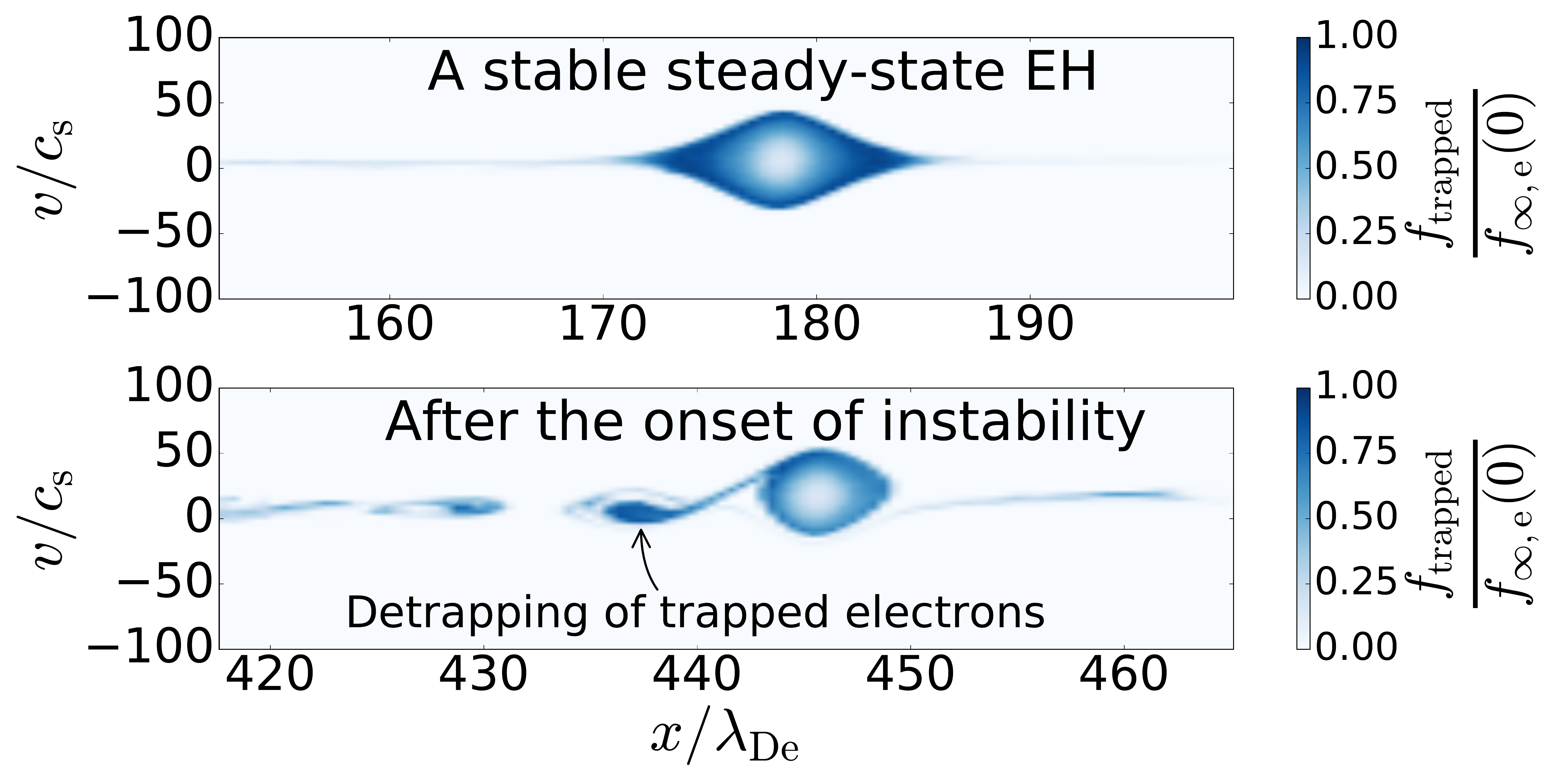}
\caption{Phase-space density of trapped electrons in our hole-tracking PIC simulation before and after the instability onset. The EH is broken into smaller pieces by this instability.}
\label{instability_detrapping}
\end{figure}

In our analysis, the trapped electrons are assumed to move with the hole potential while remaining on their trapped orbits in an oscillating hole. However, there are always shallowly trapped electrons whose slow orbits will resonate with the oscillation frequency. The fraction of these resonant particles is small and they do not much affect the linear stability analysis. Once the instability has fully developed, some resonant particles become detrapped, causing the EH to shrink in size. This is the nonlinear stage of the instability. Trapped electrons in a steady-state EH are tagged in our simulations to follow their motion \cite{Zhou2016}. The detrapping of trapped electrons by the instability is shown in Figure \ref{instability_detrapping}. It is observed that the instability can be nonlinearly saturated by the shrinking of the EH. In some other cases, the oscillation amplitude in the hole velocity grows until the EH velocity is in the close vicinity of the ion velocity. The EH is then disrupted by the ions through the mechanism described by Saeki et al. \citep{Saeki1998} The presence of a parallel electric field can slow down the EHs in the ion frame and lead to the instability. Happening in space, this instability can cause ion heating by ion Landau damping and drive anomalous resistivity \citep{Dyrud2006}. The EHs are considered to stem from phase-space instability and they are reservoirs of wave energy. The described instability provides a mechanism to couple the stored wave energy in the EH to the ion and electron plasma energy.

Our solitary wave velocity stability theory is generic and appears to apply to ion-acoustic solitons. Why are ion-acoustic solitary waves, propagating at a velocity slightly higher than $c_s$ in the ion frame, stable? $\dot{P}_e$ goes to $0$ when $|U|$ approaches $c_s$ for small wave amplitude $\psi$. Hence, for a solitary wave propagating around this velocity, solutions of our dispersion relation, $\dot{P}_i/\dot{P}_e+1=0$, exist only at very high freuqency where the short-transit-time approximation for electrons break down. So it is possible to have a stable propagation region for the solitary wave at a speed around $c_s$. For a solitary wave in this regime, $|\dot{P}_i/\dot{P}_e(\omega)|\gg 1$ for all physical frequencies of the system. The ion dynamics therefore dominates over the electron dynamics inside the solitary wave and it propagates like a Korteweg-de Vries-type ion-acoustic soliton even though an electron phase-space structure might be attached to it. 

In contrast, a solitary wave propagating much faster than $U_c$ is dominated by electron dynamics: $|\dot{P}_i/\dot{P}_e(\omega)|\ll 1$ for all physical $\omega$, and can be considered a pure electron hole. When the ion dynamics and electron dynamics are comparable inside a propagating solitary wave, our stability theory predicts that the propagation can be driven unstable by positive dynamical feedback between the two species. The velocity instability reported in this paper naturally separates these two major types of plasma electrostatic solitary waves. We will address this point in greater detail in follow up work.

\section{Conclusion} \label{sec:4}
In this paper, we've reported a new kind of instability for an EH propagating with the presence of heavier ions and we have presented the theoretical understanding behind it. An EH at low speed in the ion frame experiences unstable velocity oscillations that can be understood treating it as a holistic object. Our analytic treatment is in full agreement with the PIC simulation observations. The ``slow" EHs that are observed in space might be susceptible to this instability. We demonstrated that the velocity oscillations initially take place at the critical ion transit frequency through the hole potential and it grows quickly to a noticeable level once the speed is below the threshold. The instability happens when the electron and ion dynamics are $180$ degrees out of phase. Our discovery is a type of solitary wave instability driven by the different inertial scales of the two constituent species of the plasma. Analogies may also be found in other physical systems.

\appendix
\renewcommand{\theequation}{A-\arabic{equation}} 
\setcounter{equation}{0}
\section*{Appendix A: rate of change of contained ion momentum}
We calculate the rate of change of inertial-frame ion momentum contained in the control volume from $x_a$ to $x_b$ to the leading order in $t_{ab}\dot{U}/U$
\begin{flalign}\label{eqn:1}
\dot{P}_{ic}\,=\,&\frac{d}{dt}(\int_{x_a}^{x_b}m_in\bar{v}\,dx)\nonumber\\ 
=\,&m_i\int_{x_a}^{x_b}\frac{\partial n}{\partial t}(v+U_{\mathrm{ref}})\,dx+m_i\int_{x_a}^{x_b}n\frac{\partial (v+U_{\mathrm{ref}})}{\partial t}\,dx \nonumber\\
=\,&m_iU_{\mathrm{ref}}\int_{x_a}^{x_b}\frac{\partial n}{\partial t}\,dx+m_i\int_{x_a}^{x_b}\frac{\partial n}{\partial t}v\,dx+m_i\int_{x_a}^{x_b}n\frac{\partial v}{\partial t}\,dx+m_i\dot{U}_{\mathrm{ref}}\int_{x_a}^{x_b}n\,dx.
\end{flalign} 
$U_{\mathrm{ref}}$ is the hole velocity in the reference frame. $U_{\mathrm{ref}}$ is the characteristic of the reference frame and does not depend on $x$. We have $\dot{U}_{\mathrm{ref}}=\dot{U}$. Choose the inertial frame such that $U_{\mathrm{ref}}=0$ (the instantaneous hole rest frame). Then at $t=t_f$
\begin{flalign}\label{eqn:2}
\dot{P}_{ic}\,=\,m_i\int_{x_a}^{x_b}\frac{\partial (nv)}{\partial t}\,dx+m_i\dot{U}\int_{x_a}^{x_b}n\,dx.
\end{flalign}
We apply the continuity of an ion fluid element from $x_a$ to a position $x$ between $x_a$ and $x_b$
\begin{equation}\label{eqn:3}
n_{as_x}v_{as_x}\delta t_{as_x}\,=\,nv\delta t_{xf},
\end{equation}
where the subscript $s_x$ refers to the time $t_{s_x}$ for an ion particle at $x$ when $t=t_f$ to first enter the control volume at $x_a$, so $t_{s_x}=t_f-\int_{x_a}^{x}\frac{du}{v}$. Using the constancy of the inflow density, we express $nv$ as
\begin{flalign}\label{eqn:4}
nv\,&=\,\frac{n_{\mathrm{\infty}}v_{as_x}\delta t_{as_x}}{\delta t_{xf}}, 
\end{flalign}
with
\begin{flalign}\label{eqn:5}
\frac{\delta t_{as_x}}{\delta t_{xf}}\,\simeq\,\frac{1}{1+\displaystyle\frac{\partial t_{ax}}{\partial t_{s_x}}\bigg |_x}.
\end{flalign}
$t_{ax}=\int_{x_a}^x\frac{dx_1}{v}$ is the ion transit time between $x_a$ and $x$. The ion velocity is governed by the conservation of its first integral of motion
\begin{equation}
\frac{v^2(t(x_1,x))}{2}+\phi(x_1)=\frac{v^2(t_{s_x})}{2}-\int_{x_a}^{x_1}\dot{U}(t(x_2,x))\,dx_2.
\end{equation}
We can then give an expression for $\frac{\partial t_{ax}}{\partial t_{s_x}}|_{x}$ to its leading order
\begin{flalign}\label{eqn:8}
\frac{\partial t_{ax}}{\partial t_{s_x}}\bigg|_{x}\,&=\,\frac{\partial}{\partial t_{s_x}} \bigg(\int_{x_a}^{x}\frac{dx_1}{v}\bigg) \nonumber \\
&=\,\int_{x_a}^{x}\frac{\partial}{\partial t_{s_x}}\bigg|_{x_1}\bigg(\frac{1}{v}\bigg)\, dx_1 \nonumber \\
&=\,\int_{x_a}^{x}-\frac{1}{v^3}\frac{\partial (v^2/2)}{\partial t_{s_x}}\bigg|_{x_1}\,dx_1 \nonumber \\
&=\, \int_{x_a}^{x}\frac{1}{v^3}\bigg[v(t_{s_x})\dot{U}(t_{s_x})+\int_{x_a}^{x_1}\frac{\partial \dot{U}}{\partial t_{s_x}}\bigg|_{x_2}\,dx_2\bigg]\,dx_1\nonumber \\
&\simeq\,\int_{x_a}^{x}\frac{v(t_{s_x})}{v^3}\bigg[\dot{U}(t_{s_x})+\frac{1}{v(t_{s_x})}\int_{x_a}^{x_1}\ddot{U}(t(x_2,x))\,dx_2\bigg]\,dx_1.
\end{flalign}
We used interchangeably $dt$ and $dt_{sx}$ as $t(x_2,x)=t_f-\int_{x_2}^{x}\frac{du}{v}$ and $t(x)=t_{s_x}+\int_{x_a}^{x}\frac{dx_1}{v_0(x_1)+v_1(x_1,t_{s_x})}$ so that $dt=dt_{s_x}(1+\mathcal{O}(t_{ab}\dot{U}/U))$ for $x$ fixed. The first term of $\dot{P}_{ic}$ in Eq. \eqref{eqn:2} can be expressed as
\begin{flalign}\label{eqn:9}
m_i\int_{x_a}^{x_b}\frac{\partial (nv)}{\partial t}\bigg|_{x}\,dx\,&\simeq\,m_i\int_{x_a}^{x_b}\frac{\partial (nv)}{\partial t_{s_x}}\bigg|_{x}\,dx \nonumber \\
&=\,m_i\int_{x_a}^{x_b}\frac{\partial}{\partial t_{s_x}}\bigg|_{x}\bigg(n_{\infty}v(t_{s_x})\frac{\delta t_{as_x}}{\delta t_{xf}}\bigg)\,dx \nonumber \\
&=\,\underbrace{-m_in_{\infty}\int_{x_a}^{x_b}\dot{U}(t_{s_x})\frac{\delta t_{as_x}}{\delta t_{xf}}\,dx}_\text{\textbf{I}}+\underbrace{m_in_{\infty}\int_{x_a}^{x_b}v(t_{s_x})\frac{\partial}{\partial t_{s_x}}\bigg|_{x}\bigg(\frac{\delta t_{as_x}}{\delta t_{xf}}\bigg)\,dx}_\text{\textbf{II}}.
\end{flalign}
Combining Eqn. \eqref{eqn:8} and Eqn. \eqref{eqn:5}, we have an expression for $\frac{\delta t_{as_x}}{\delta t_{xf}}$
\begin{flalign}\label{eqn:10}
\frac{\delta t_{as_x}}{\delta t_{xf}}\,&\simeq\,\frac{1}{1+\displaystyle \int_{x_a}^{x}\frac{v(t_{s_x})}{v^3}\bigg[\dot{U}(t_{s_x})+\frac{1}{v(t_{s_x})}\int_{x_a}^{x_1}\ddot{U}(t(x_2,x))\,dx_2\bigg]\,dx_1} \nonumber \\
&=\,1+\mathcal{O}\bigg(\frac{t_{ab} \dot{U}}{U}\bigg).
\end{flalign}
Therefore, to the lowest order, the term \textbf{I} in Eq. \eqref{eqn:9} is
\begin{flalign}\label{eqn:11}
-m_in_{\infty}\int_{x_a}^{x_b}\dot{U}(t_{s_x})\,dx.
\end{flalign}
To calculate the term \textbf{II} in Eq. \eqref{eqn:9}, we need
\begin{flalign}\label{eqn:12}
\frac{\partial}{\partial t_{s_x}}\bigg|_{x}\bigg(\frac{\delta t_{as_x}}{\delta t_{xf}}\bigg)\,=\,-\frac{1}{\displaystyle \bigg(1+\frac{\partial t_{ax}}{\partial t_{s_x}}\bigg|_{x}\bigg)^2}\frac{\partial^2 t_{ax}}{\partial t_{s_x}^2}\bigg|_{x}.
\end{flalign}
We use the results from Eq. \eqref{eqn:8} to get
\begin{flalign}\label{eqn:13}
\frac{\partial^2 t_{ax}}{\partial t_{s_x}^2}\bigg|_{x}\,&=\,\int_{x_a}^{x}\frac{\partial}{\partial t_{s_x}}\bigg(-\frac{1}{v^3}\frac{\partial (v^2/2)}{\partial t_{s_x}}\bigg)\,dx_1 \nonumber \\
&\simeq\,\int_{x_a}^{x}\frac{1}{v^3}\frac{\partial}{\partial t_{s_x}}\bigg(v(t_{s_x})\dot{U}(t_{s_x})+\int_{x_a}^{x_1}\ddot{U}(t(x_2,x))\,dx_2\bigg)\,dx_1\,\,\,(\mathrm{1^{st}\,\,order}) \nonumber \\
&\simeq\,\int_{x_a}^{x}\frac{v(t_{s_x})}{v^3}\bigg(\ddot{U}(t_{s_x})+\frac{1}{v(t_{s_x})}\int_{x_a}^{x_1}\dddot{U}(t(x_2,x))\,dx_2\bigg)\,dx_1.
\end{flalign}
Thus, the term \textbf{II} in Eq. \eqref{eqn:9} can be written as
\begin{flalign}\label{eqn:14}
-m_in_{\infty}\int_{x_a}^{x_b}v^2(t_{s_x})\int_{x_a}^{x}\frac{1}{v^3}\bigg(\ddot{U}(t_{s_x})+\frac{1}{v(t_{s_x})}\int_{x_a}^{x_1}\dddot{U}(t(x_2,x))\,dx_2\bigg)\,dx_1\,dx.
\end{flalign}
Combining everything above, we have to the leading order
\begin{flalign}\label{eqn:15}
\dot{P}_{ic}\,=\,&-m_in_{\infty}\int_{x_a}^{x_b}\dot{U}(t_{s_x})\,dx-m_in_{\infty}\dot{U}(t_f)\int_{x_a}^{x_b}\frac{U}{v}\,dx \nonumber \\
&-m_in_{\infty}\int_{x_a}^{x_b}v^2(t_{s_x})\int_{x_a}^{x}\frac{1}{v^3}\bigg(\ddot{U}(t_{s_x})+\frac{1}{v(t_{s_x})}\int_{x_a}^{x_1}\dddot{U}(t(x_2,x))\,dx_2\bigg)\,dx_1\,dx.
\end{flalign}
Now we apply integration by parts to the last term in Eq. \eqref{eqn:15} using $\frac{\partial t(x_2,x)}{\partial x_2}|_{x}=\frac{1}{v(t(x_2,x))}$
\begin{flalign}\label{eqn:17}
\ddot{U}(t_{s_x})+\frac{1}{v(t_{s_x})}\int_{x_a}^{x_1}\dddot{U}(t(x_2,x))\,dx_2\,&=\,\ddot{U}(t_f-\int_{x_a}^{x}\frac{du}{v})+\frac{1}{v(t_{s_x})}\int_{x_a}^{x_1}\dddot{U}(t_f-\int_{x_2}^{x}\frac{du}{v})\frac{1}{v}v\,dx_2 \nonumber \\
&=\,\ddot{U}(t_{s_x})+\frac{1}{v(t_{s_x})}\bigg(\bigg[v\ddot{U}\bigg]_{x_a}^{x_1}-\int_{x_a}^{x_1}\ddot{U}\frac{\dot{v}}{v}\,dx_2\bigg) \nonumber \\
&=\,\ddot{U}(t(x_1,x))\frac{v}{v(t_{s_x})}-\frac{1}{v(t_{s_x})}\int_{x_a}^{x_1}\frac{\ddot{U}(t(x_2,x))\dot{v}}{v}\,dx_2
\end{flalign}
Plug this expression in Eq. \eqref{eqn:15} and remember that $v(t_{s_x})\simeq-U$, we have
\begin{flalign}\label{eqn:18}
\dot{P}_{ic}\,=\,&-m_in_{\infty}\int_{x_a}^{x_b}\dot{U}(t_{s_x})\,dx-m_in_{\infty}\dot{U}(t_f)\int_{x_a}^{x_b}\frac{U}{v}\,dx-m_in_{\infty}\int_{x_a}^{x_b}v(t_{s_x})\int_{x_a}^{x}\frac{\ddot{U}(t(x_1,x))}{v^2}\,dx_1\,dx \nonumber \\
&+m_in_{\infty}\int_{x_a}^{x_b}v(t_{s_x})\int_{x_a}^{x}\frac{1}{v^3}\int_{x_a}^{x_1}\frac{\ddot{U}(t(x_2,x))\dot{v}}{v}\,dx_2\,dx_1\,dx,
\end{flalign}
where
\begin{flalign}\label{eqn:19}
\int_{x_a}^x\frac{\ddot{U}(t(x_1,x))}{v^2}\,dx_1\,&=\,\bigg[\frac{1}{v}\dot{U}(t_f-\int_{x_1}^{x}\frac{du}{v})\bigg]_{x_a}^x+\int_{x_a}^x\frac{\dot{U}(t(x_1,x))\dot{v}}{v^3}\,dx_1 \nonumber \\
&=\,\frac{\dot{U}(t_f)}{v}-\frac{\dot{U}(t_{s_x})}{v(t_{s_x})}+\int_{x_a}^x\frac{\dot{U}(t(x_1,x))\dot{v}}{v^3}\,dx_1
\end{flalign}
and
\begin{flalign}\label{eqn:20}
\int_{x_a}^{x_1}\frac{\ddot{U}(t(x_2,x))\dot{v}}{v}\,dx_2\,&=\,\bigg[\dot{U}\dot{v}(t(x_2,x))\bigg]_{x_a}^{x_1}-\int_{x_a}^{x_1}\frac{\dot{U}\ddot{v}}{v}(t(x_2,x))\,dx_2 \nonumber \\
&\simeq\,\dot{U}\dot{v}(t(x_1,x))-\int_{x_a}^{x_1}\frac{\dot{U}\ddot{v}}{v}(t(x_2,x))\,dx_2\,\,\,(\mathrm{1^{st}\,\,order}).
\end{flalign}
Simplify Eq. \eqref{eqn:18} by replacing the $\ddot{U}$ terms using Eqs. \eqref{eqn:19} \eqref{eqn:20}. Most terms cancel out to the relevant order and we are left with only one leading order term
\begin{flalign}\label{eqn:21}
\dot{P}_{ic}\,=\,&-m_in_{\infty}\int_{x_a}^{x_b}v(t_{s_x})\int_{x_a}^{x}\frac{1}{v^3}\int_{x_a}^{x_1}\frac{\dot{U}\ddot{v}}{v}(t(x_2,x))\,dx_2\,dx_1\,dx \nonumber \\
\simeq\,&\,-m_in_{\infty}\int_{x_a}^{x_b}\int_{x_a}^{x}\frac{U}{v_0^3(x_1)}\int_{x_a}^{x_1}\dot{U}(t(x_2,x))\phi''(x_2)\,dx_2\,dx_1\,dx,
\end{flalign}
where we used the equation of motion for a single ion particle
\begin{flalign}\label{eqn:22}
\dot{v}+\phi'\,=\,-\dot{U}.
\end{flalign}
We combine $\dot{P}_{io}$ and $\dot{P}_{ic}$ to get $\dot{P}_i$
\begin{flalign}\label{eqn:23}
\dot{P}_i\,=\,\dot{P}_{io}+\dot{P}_{ic}.
\end{flalign}
\renewcommand{\theequation}{B-\arabic{equation}} 
\setcounter{equation}{0}
\section*{Appendix B: Leading order expansion of $\dot{P}_i/\dot{P}_e$ in $\psi/U^2$}
\subsection*{First order in $\psi/U^2$}
The full expression of $\dot{P}_i$ is
\begin{equation}\label{eqn:P_i}
\begin{aligned}
\dot{P}_i\,=&\,n_{\infty}m_i\dot{U}\bigg [\displaystyle \int_{x_a}^{x_b}\left(-1-2\displaystyle \frac{U}{v_0(x)}-\displaystyle \frac{U^2}{v_0^2(x)}\right)\exp\left(i\omega\displaystyle \int_{x}^{x_b}\displaystyle \frac{dx_3}{v_0(x_3)}\right)\,dx\\
&\hspace{1cm}-\displaystyle \int_{x_a}^{x_b} \displaystyle \frac{U^2}{v_0^3(x)} \displaystyle \int_{x_a}^{x}\displaystyle \frac{\phi'(x_1)}{v_0(x_1)}\exp\left(i\omega\displaystyle \int_{x_1}^{x_b}\displaystyle \frac{dx_3}{v_0(x_3)}\right)\,dx_1\,dx\\ 
&\hspace{1cm}-\displaystyle \int_{x_a}^{x_b}\displaystyle \int_{x_a}^{x}\displaystyle \frac{U}{v_0^3(x_1)}\displaystyle \int_{x_a}^{x_1}\exp\left(i\omega\displaystyle \int_{x_2}^{x}\displaystyle \frac{dx_3}{v_0(x_3)}\right)\phi''(x_2)\,dx_2\,dx_1\,dx\bigg ].
\end{aligned}
\end{equation}
We will begin by expanding the expression of $\dot{P}_i$ to the first order in $\psi/U^2$. Recall that the equilibrium ion velocity in the hole frame $v_0$ is
\begin{equation}
\begin{aligned}
v_0(x)\,&=\,-\frac{U}{|U|}\sqrt{U^2-2\phi(x)}\\
&\simeq\,-U(1-\frac{\psi}{U^2}\tilde{\phi}(x))
\end{aligned}
\end{equation}
And the phase term inside the integral sign of Eqn. \eqref{eqn:P_i} is
\begin{equation}
\begin{aligned}
\exp(i\omega\int_{x}^{x_b}\frac{dx_3}{v_0(x_3)})\,\simeq\,\exp(i\frac{\omega(x-x_b)}{U})(1-\frac{i\omega}{U}\frac{\psi}{U^2}\int_{x}^{x_b}\tilde{\phi}(x_3)\,dx_3).
\end{aligned}
\end{equation}
The term inside the first integral can be rewritten taking $v_0\simeq-U$ as
\begin{equation}
-1-2\displaystyle \frac{U}{v_0(x)}-\displaystyle \frac{U^2}{v_0^2(x)}\,=\,-\left(\frac{U}{v_0(x)}+1\right)^2\,\simeq\,-\frac{\psi^2}{U^4}\tilde{\phi}(x)^2,
\end{equation}
which is already a second order term. The only first order contributions come from the last two integrals. To the first order in $\psi/U^2$, $\dot{P}_i$ can be written as
\begin{equation}\label{eqn:first_order}
\begin{aligned}
\dot{P}_i\,\simeq\,&n_{\infty}m_i\dot{U}\frac{\psi}{U^2}\bigg[-\int_{x_a}^{x_b}\int_{x_a}^{x}\tilde{\phi}'(x_1)\exp\left(i\frac{\omega}{U}(x_1-x_b)\right)\,dx_1\,dx\\
&\hspace{1cm}+\int_{x_a}^{x_b}\int_{x_a}^{x}\int_{x_a}^{x_1}\tilde{\phi}''(x_2)\exp\left(i\frac{\omega}{U}(x_2-x)\right)\,dx_2\,dx_1\,dx \bigg].
\end{aligned}
\end{equation}
We first apply integration by parts to the triple integral term in Eqn. \eqref{eqn:first_order}
\begin{equation}\label{eqn:B_6}
\begin{aligned}
&\int_{x_a}^{x_b}\int_{x_a}^{x}\int_{x_a}^{x_1}\tilde{\phi}''(x_2)\exp\left(i\frac{\omega}{U}(x_2-x)\right)\,dx_2\,dx_1\,dx\,=\,\int_{x_a}^{x_b}\int_{x_a}^{x}\tilde{\phi}'(x_1)\exp\left(i\frac{\omega}{U}(x_1-x)\right)\,dx_1\,dx\\
&\hspace{1cm}-\int_{x_a}^{x_b}\int_{x_a}^{x}\int_{x_a}^{x_1}\tilde{\phi}'(x_2)\exp\left(i\frac{\omega}{U}(x_2-x)\right)\left(\frac{i\omega}{U}\right)\,dx_2\,dx_1\,dx.
\end{aligned}
\end{equation}
We interchange the order of integration to integrate the triple integral. The integration domain of this triple integral is
\begin{equation}
\mathcal{D}\,=\,\left\{(x,x_1,x_2)\in\mathds{R}^3|x_a\leq x_2\leq x_1 \leq x\leq x_b \right \}.
\end{equation}
We define an indicator function $\mathds{1}_{\mathcal{D}}$ associated with this measurable subset of $\mathds{R}^3$ and we have 
\begin{equation}\label{eqn:B_8}
\begin{aligned}
&\int_{x_a}^{x_b}\int_{x_a}^{x}\int_{x_a}^{x_1}\tilde{\phi}'(x_2)\exp\left(i\frac{\omega}{U}(x_2-x)\right)\left(\frac{i\omega}{U}\right)\,dx_2\,dx_1\,dx \\
=\,&\int_{x_a}^{x_b}\int_{x_a}^{x_b}\int_{x_a}^{x_b}\tilde{\phi}'(x_2)\exp\left(i\frac{\omega}{U}(x_2-x)\right)\left(\frac{i\omega}{U}\right)\mathds{1}_{\mathcal{D}} \,dx_2\,dx_1\,dx\\
=\,&\int_{x_a}^{x_b}\int_{x_a}^{x_b}\int_{x_a}^{x_b}\tilde{\phi}'(x_2)\exp\left(i\frac{\omega}{U}(x_2-x)\right)\left(\frac{i\omega}{U}\right)\mathds{1}_{\mathcal{D}} \,dx\,dx_2\,dx_1\;\mathrm{(Interchange\:integration\:order)}\\
=\,&\int_{x_a}^{x_b}\int_{x_a}^{x_1}\left[\tilde{\phi}'(x_2)\exp\left(i\frac{\omega}{U}(x_2-x)\right)\left(\frac{i\omega}{U}\right)\left(\frac{U}{-i\omega}\right)\right]_{x=x_1}^{x=x_b}\,dx_2\,dx_1\\
=\,&\int_{x_a}^{x_b}\int_{x_a}^{x_1}\tilde{\phi}'(x_2)\exp\left(i\frac{\omega}{U}(x_2-x_1)\right)\,dx_2\,dx_1-\int_{x_a}^{x_b}\int_{x_a}^{x_1}\tilde{\phi}'(x_2)\exp\left(i\frac{\omega}{U}(x_2-x_b)\right)\,dx_2\,dx_1.
\end{aligned}
\end{equation}
Interchanging the order of integration is the mathematical technique we use throughout this derivation to deal with multiple integrals. Since the first term in Eqn. \eqref{eqn:B_8} cancels the first term in Eqn. \eqref{eqn:B_6} and 
\begin{equation}
\int_{x_a}^{x_b}\int_{x_a}^{x}\int_{x_a}^{x_1}\tilde{\phi}''(x_2)\exp\left(i\frac{\omega}{U}(x_2-x)\right)\,dx_2\,dx_1\,dx\,=\,\int_{x_a}^{x_b}\int_{x_a}^{x_1}\tilde{\phi}'(x_2)\exp\left(i\frac{\omega}{U}(x_2-x_b)\right)\,dx_2\,dx_1.
\end{equation}
Using this result in Eqn. \eqref{eqn:first_order} and we get the exact cancellation of the two first-order terms. Therefore, we need to push the expansion to the next order, which is $\psi^2/U^4$.

\subsection*{Second order in $\psi/U^2$}
First, we identify all the terms that are of order $\psi^2/U^4$ in the expansion of $\dot{P}_i$. We examine each term in Eqn. \eqref{eqn:P_i}. We have for the first term
\begin{equation}\label{eqn:1st_term}
\begin{aligned}
&\int_{x_a}^{x_b}\left(-1-2\displaystyle \frac{U}{v_0(x)}-\displaystyle \frac{U^2}{v_0^2(x)}\right)\exp\left(i\omega\displaystyle \int_{x}^{x_b}\displaystyle \frac{dx_3}{v_0(x_3)}\right)\,dx\\
=\,&\int_{x_a}^{x_b}-\left(1+\frac{U}{v_0(x)}\right)^2\exp\left(i\omega\displaystyle \int_{x}^{x_b}\displaystyle \frac{dx_3}{v_0(x_3)}\right)\,dx\\
\simeq\,&-\frac{\psi^2}{U^4}\int_{x_a}^{x_b}\tilde{\phi}(x)^2\exp\left(i\frac{\omega}{U}(x-x_b)\right)\,dx.
\end{aligned}
\end{equation}
The second term will give contributions both from $v_0$ and the phase term
\begin{equation}\label{eqn:2nd_term}
\begin{aligned}
&-\displaystyle \int_{x_a}^{x_b} \displaystyle \frac{U^2}{v_0^3(x)} \displaystyle \int_{x_a}^{x}\displaystyle \frac{\phi'(x_1)}{v_0(x_1)}\exp\left(i\omega\displaystyle \int_{x_1}^{x_b}\displaystyle \frac{dx_3}{v_0(x_3)}\right)\,dx_1\,dx\\
\simeq\,&-\frac{\psi}{U^2}\int_{x_a}^{x_b}\int_{x_a}^{x}\tilde{\phi}'(x_1)\exp\left(i\frac{\omega}{U}(x_1-x_b)\right)\,dx_1\,dx\;\;(\mathrm{First\;order})\\
&+\frac{\psi^2}{U^4}\bigg \{-\int_{x_a}^{x_b}\int_{x_a}^{x}\left(3\tilde{\phi}(x)\tilde{\phi}'(x_1)+\tilde{\phi}'(x_1)\tilde{\phi}(x_1)\right)\exp\left(i\frac{\omega}{U}(x_1-x_b)\right)\,dx_1\,dx\\
&\hspace{1cm}+\int_{x_a}^{x_b}\int_{x_a}^{x}\int_{x_1}^{x_b}\tilde{\phi}(x_3)\tilde{\phi}'(x_1)\exp\left(i\frac{\omega}{U}(x_1-x_b)\right)\left(\frac{i\omega}{U}\right)\,dx_3\,dx_1\,dx\bigg\}.
\end{aligned}
\end{equation}
The third term can be expanded as
\begin{equation}\label{eqn:3rd_term}
\begin{aligned}
&-\displaystyle \int_{x_a}^{x_b}\displaystyle \int_{x_a}^{x}\displaystyle \frac{U}{v_0^3(x_1)}\displaystyle \int_{x_a}^{x_1}\exp\left(i\omega\displaystyle \int_{x_2}^{x}\displaystyle \frac{dx_3}{v_0(x_3)}\right)\phi''(x_2)\,dx_2\,dx_1\,dx\\
\simeq\,&\frac{\psi}{U^2}\int_{x_a}^{x_b}\int_{x_a}^{x}\int_{x_a}^{x_1}\tilde{\phi}''(x_2)\exp\left(i\frac{\omega}{U}(x_2-x)\right)\,dx_2\,dx_1\,dx\;\;(\mathrm{First\;order})\\
&+\frac{\psi^2}{U^4}\bigg\{\int_{x_a}^{x_b}\int_{x_a}^{x}\int_{x_a}^{x_1}3\tilde{\phi}(x_1)\tilde{\phi}''(x_2)\exp\left(i\frac{\omega}{U}(x_2-x)\right)\,dx_2\,dx_1\,dx\\
&\hspace{1cm}+\int_{x_a}^{x_b}\int_{x_a}^{x}\int_{x_a}^{x_1}\int_{x_2}^{x}\tilde{\phi}''(x_2)\tilde{\phi}(x_3)\exp\left(i\frac{\omega}{U}(x_2-x)\right)\left(-i\frac{\omega}{U}\right)\,dx_3\,dx_2\,dx_1\,dx\bigg \}.
\end{aligned}
\end{equation}
We will deal with these integrals one by one, starting from the simpler double integrals and moving our way to the quadruple integral. The idea is to use integration by parts and interchanging the order of integration to simplify them as much as possible as we showed previously. In the end, we should have simpler integral expressions involving only $\tilde{\phi}$ and not its derivatives. Let's start with the double integral that is multiplying the second order term in Eqn. \eqref{eqn:2nd_term}.
\begin{equation}\label{eqn:final_1}
\begin{aligned}
&-\int_{x_a}^{x_b}\int_{x_a}^{x}\left(3\tilde{\phi}(x)\tilde{\phi}'(x_1)+\tilde{\phi}'(x_1)\tilde{\phi}(x_1)\right)\exp\left(i\frac{\omega}{U}(x_1-x_b)\right)\,dx_1\,dx\\
=&\,-\int_{x_a}^{x_b}\int_{x_a}^{x}3\tilde{\phi}(x)\exp\left(i\frac{\omega}{U}(x_1-x_b)\right)\,d\tilde{\phi}(x_1)\,dx-\int_{x_a}^{x_b}\int_{x_a}^{x}\exp\left(i\frac{\omega}{U}(x_1-x_b)\right)\,d(\tilde{\phi}(x_1)^2/2)\,dx\\
=&\,-\frac{7}{2}\int_{x_a}^{x_b}\tilde{\phi}(x)^2\exp\left(i\frac{\omega}{U}(x-x_b)\right)\,dx+3\left(\frac{i\omega}{U}\right)\int_{x_a}^{x_b}\int_{x_a}^{x}\tilde{\phi}(x)\tilde{\phi}(x_1)\exp\left(i\frac{\omega}{U}(x_1-x_b)\right)\,dx_1\,dx\\
&\hspace{1cm}+\left(\frac{i\omega}{U}\right)\int_{x_a}^{x_b}\int_{x_a}^{x}\frac{\tilde{\phi}(x_1)^2}{2}\exp\left(i\frac{\omega}{U}(x_1-x_b)\right)\,dx_1\,dx\\
=&\,-\frac{7}{2}\int_{x_a}^{x_b}\tilde{\phi}(x)^2\exp\left(i\frac{\omega}{U}(x-x_b)\right)\,dx+3\left(\frac{i\omega}{U}\right)\int_{x_a}^{x_b}\int_{x_a}^{x}\tilde{\phi}(x)\tilde{\phi}(x_1)\exp\left(i\frac{\omega}{U}(x_1-x_b)\right)\,dx_1\,dx\\
&\hspace{1cm}+\frac{1}{2}\left(\frac{i\omega}{U}\right)\int_{x_a}^{x_b}\int_{x_1}^{x_b}\tilde{\phi}(x_1)^2\exp\left(i\frac{\omega}{U}(x_1-x_b)\right)\,dx\,dx_1\\
=&\,-\frac{7}{2}\int_{x_a}^{x_b}\tilde{\phi}(x)^2\exp\left(i\frac{\omega}{U}(x-x_b)\right)\,dx+3\left(\frac{i\omega}{U}\right)\int_{x_a}^{x_b}\int_{x_a}^{x}\tilde{\phi}(x)\tilde{\phi}(x_1)\exp\left(i\frac{\omega}{U}(x_1-x_b)\right)\,dx_1\,dx\\
&\hspace{1cm}+\frac{1}{2}\left(\frac{i\omega}{U}\right)\int_{x_a}^{x_b}\tilde{\phi}(x_1)^2\exp\left(i\frac{\omega}{U}(x_1-x_b)\right)(x_b-x_1)\,dx_1.
\end{aligned}
\end{equation}
The triple integral in Eqn. \eqref{eqn:2nd_term} is 
\begin{equation}\label{eqn:final_2}
\begin{aligned}
&\int_{x_a}^{x_b}\int_{x_a}^{x}\int_{x_1}^{x_b}\tilde{\phi}(x_3)\tilde{\phi}'(x_1)\exp\left(i\frac{\omega}{U}(x_1-x_b)\right)\left(\frac{i\omega}{U}\right)\,dx_3\,dx_1\,dx\\
=\,&\int_{x_a}^{x_b}\int_{x_a}^{x_3}\int_{x_1}^{x_b}\tilde{\phi}(x_3)\tilde{\phi}'(x_1)\exp\left(i\frac{\omega}{U}(x_1-x_b)\right)\left(\frac{i\omega}{U}\right)\,dx\,dx_1\,dx_3\\
=\,&\int_{x_a}^{x_b}\int_{x_a}^{x_3}\tilde{\phi}(x_3)\exp\left(i\frac{\omega}{U}(x_1-x_b)\right)(x_b-x_1)\left(\frac{i\omega}{U}\right)\,d\phi(x_1)\,dx_3\\
=\,&\left(\frac{i\omega}{U}\right)\int_{x_a}^{x_b}\tilde{\phi}(x_3)^2(x_b-x_3)\exp\left(i\frac{\omega}{U}(x_3-x_b)\right)\,dx_3\\
&\hspace{1cm}+\left(\frac{i\omega}{U}\right)\int_{x_a}^{x_b}\int_{x_a}^{x_3}\tilde{\phi}(x_3)\tilde{\phi}(x_1)\exp\left(i\frac{\omega}{U}(x_1-x_b)\right)\,dx_1\,dx_3\\
&\hspace{1cm}-\left(\frac{i\omega}{U}\right)^2\int_{x_a}^{x_b}\int_{x_a}^{x_3}\tilde{\phi}(x_3)\tilde{\phi}(x_1)(x_b-x_1)\exp\left(i\frac{\omega}{U}(x_1-x_b)\right)\,dx_1\,dx_3.
\end{aligned}
\end{equation}
Now we simplify the triple integral in Eqn. \eqref{eqn:3rd_term} using twice integration by parts and interchanging the order of integration
\begin{equation}\label{eqn:final_3}
\begin{aligned}
&\int_{x_a}^{x_b}\int_{x_a}^{x}\int_{x_a}^{x_1}3\tilde{\phi}(x_1)\tilde{\phi}''(x_2)\exp\left(i\frac{\omega}{U}(x_2-x)\right)\,dx_2\,dx_1\,dx\\
=\,&3\int_{x_a}^{x_b}\int_{x_a}^{x}\tilde{\phi}(x_1)\tilde{\phi}'(x_1)\exp\left(i\frac{\omega}{U}(x_1-x)\right)\,dx_1\,dx\\
&\hspace{1cm}-3\int_{x_a}^{x_b}\int_{x_a}^{x}\int_{x_a}^{x_1}\tilde{\phi}(x_1)\tilde{\phi}'(x_2)\exp\left(i\frac{\omega}{U}(x_2-x)\right)\left(\frac{i\omega}{U}\right)\,dx_2\,dx_1\,dx\\
=\,&\frac{3}{2}\int_{x_a}^{x_b}\tilde{\phi}(x)^2\,dx-\frac{9}{2}\int_{x_a}^{x_b}\int_{x_a}^x\tilde{\phi}(x_1)^2\exp\left(i\frac{\omega}{U}(x_1-x)\right)\left(\frac{i\omega}{U}\right)\,dx_1\,dx\\
&\hspace{1cm}+3\int_{x_a}^{x_b}\int_{x_a}^{x_1}\int_{x_1}^{x_b}\tilde{\phi}(x_1)\tilde{\phi}(x_2)\exp\left(i\frac{\omega}{U}(x_2-x)\right)\left(\frac{i\omega}{U}\right)^2\,dx\,dx_2\,dx_1\\
=\,&\frac{3}{2}\int_{x_a}^{x_b}\tilde{\phi}(x)^2\,dx-\frac{9}{2}\left(\frac{i\omega}{U}\right)\int_{x_a}^{x_b}\bigg[\tilde{\phi}(x_1)^2\exp\left(i\frac{\omega}{U}(x_1-x)\right)\left(\frac{-U}{i\omega}\right)\bigg]_{x=x_1}^{x=x_b}\,dx_1\\
&\hspace{1cm}-3\left(\frac{i\omega}{U}\right)\int_{x_a}^{x_b}\int_{x_a}^{x_1}\tilde{\phi}(x_1)\tilde{\phi}(x_2)\exp\left(i\frac{\omega}{U}(x_2-x_b)\right)\,dx_2\,dx_1\\
&\hspace{1cm}+3\left(\frac{i\omega}{U}\right)\int_{x_a}^{x_b}\int_{x_a}^{x_1}\tilde{\phi}(x_1)\tilde{\phi}(x_2)\exp\left(i\frac{\omega}{U}(x_2-x_1)\right)\,dx_2\,dx_1\\
=\,&-3\int_{x_a}^{x_b}\tilde{\phi}(x)^2\,dx+\frac{9}{2}\int_{x_a}^{x_b}\tilde{\phi}(x_1)^2\exp\left(i\frac{\omega}{U}(x_1-x_b)\right)\,dx_1\\
&\hspace{1cm}-3\left(\frac{i\omega}{U}\right)\int_{x_a}^{x_b}\int_{x_a}^{x_1}\tilde{\phi}(x_1)\tilde{\phi}(x_2)\exp\left(i\frac{\omega}{U}(x_2-x_b)\right)\,dx_2\,dx_1\\
&\hspace{1cm}+3\left(\frac{i\omega}{U}\right)\int_{x_a}^{x_b}\int_{x_a}^{x_1}\tilde{\phi}(x_1)\tilde{\phi}(x_2)\exp\left(i\frac{\omega}{U}(x_2-x_1)\right)\,dx_2\,dx_1.
\end{aligned}
\end{equation}
Now we deal with the only term left, which is the quadruple integral in the Eqn. \eqref{eqn:3rd_term}. We will try to relate it to the integrals we've already calculated. Observing the integrand, we notice that it is easy to integrate with respect to the variable $x_1$. We interchange the order of integration as we did before
\begin{equation}
\begin{aligned}
&\int_{x_a}^{x_b}\int_{x_a}^{x}\int_{x_a}^{x_1}\int_{x_2}^{x}\tilde{\phi}''(x_2)\tilde{\phi}(x_3)\exp\left(i\frac{\omega}{U}(x_2-x)\right)\left(-i\frac{\omega}{U}\right)\,dx_3\,dx_2\,dx_1\,dx\\
=\,&\int_{x_a}^{x_b}\int_{x_a}^{x}\int_{x_2}^{x}\int_{x_2}^{x}\tilde{\phi}''(x_2)\tilde{\phi}(x_3)\exp\left(i\frac{\omega}{U}(x_2-x)\right)\left(-i\frac{\omega}{U}\right)\,dx_1\,dx_3\,dx_2\,dx\\
=\,&\int_{x_a}^{x_b}\int_{x_a}^{x}\int_{x_2}^{x}\tilde{\phi}''(x_2)\tilde{\phi}(x_3)\exp\left(i\frac{\omega}{U}(x_2-x)\right)\left(\frac{i\omega}{U}\right)(x_2-x)\,dx_3\,dx_2\,dx.
\end{aligned}
\end{equation}
Notice that the above integral can be expressed with the derivative w.r.t. $(i\omega/U)$ of another integral that we've calculated previously in the beginning of Eqn. \eqref{eqn:final_3}
\begin{equation}
\begin{aligned}
&\int_{x_a}^{x_b}\int_{x_a}^{x}\int_{x_2}^{x}\tilde{\phi}''(x_2)\tilde{\phi}(x_3)\exp\left(i\frac{\omega}{U}(x_2-x)\right)\left(\frac{i\omega}{U}\right)(x_2-x)\,dx_3\,dx_2\,dx\\
=\,&\left(\frac{i\omega}{U}\right)\frac{d}{d\left(i\omega/U\right)}\bigg[\int_{x_a}^{x_b}\int_{x_a}^{x}\int_{x_2}^{x}\tilde{\phi}''(x_2)\tilde{\phi}(x_3)\exp\left(i\frac{\omega}{U}(x_2-x)\right)\,dx_3\,dx_2\,dx\bigg]\\
=\,&\left(\frac{i\omega}{U}\right)\frac{d}{d\left(i\omega/U\right)}\bigg[\int_{x_a}^{x_b}\int_{x_a}^{x}\int_{x_a}^{x_3}\tilde{\phi}''(x_2)\tilde{\phi}(x_3)\exp\left(i\frac{\omega}{U}(x_2-x)\right)\,dx_2\,dx_3\,dx\bigg].
\end{aligned}
\end{equation}
Therefore, the quadruple integral can be simplified using the final result of Eqn. \eqref{eqn:final_3}
\begin{equation}\label{eqn:final_4}
\begin{aligned}
&\int_{x_a}^{x_b}\int_{x_a}^{x}\int_{x_a}^{x_1}\int_{x_2}^{x}\tilde{\phi}''(x_2)\tilde{\phi}(x_3)\exp\left(i\frac{\omega}{U}(x_2-x)\right)\left(-i\frac{\omega}{U}\right)\,dx_3\,dx_2\,dx_1\,dx\\
=\,&\left(\frac{i\omega}{U}\right)\frac{d}{d\left(i\omega/U\right)}\bigg[\int_{x_a}^{x_b}\int_{x_a}^{x}\int_{x_a}^{x_3}\tilde{\phi}''(x_2)\tilde{\phi}(x_3)\exp\left(i\frac{\omega}{U}(x_2-x)\right)\,dx_2\,dx_3\,dx\bigg]\\
=\,&\left(\frac{i\omega}{U}\right)\frac{d}{d\left(i\omega/U\right)}\bigg[-\int_{x_a}^{x_b}\tilde{\phi}(x)^2\,dx+\frac{3}{2}\int_{x_a}^{x_b}\tilde{\phi}(x_1)^2\exp\left(i\frac{\omega}{U}(x_1-x_b)\right)\,dx_1\\
&\hspace{1cm}-\left(\frac{i\omega}{U}\right)\int_{x_a}^{x_b}\int_{x_a}^{x_1}\tilde{\phi}(x_1)\tilde{\phi}(x_2)\exp\left(i\frac{\omega}{U}(x_2-x_b)\right)\,dx_2\,dx_1\\
&\hspace{1cm}+\left(\frac{i\omega}{U}\right)\int_{x_a}^{x_b}\int_{x_a}^{x_1}\tilde{\phi}(x_1)\tilde{\phi}(x_2)\exp\left(i\frac{\omega}{U}(x_2-x_1)\right)\,dx_2\,dx_1\bigg]\\
=\,&\frac{3}{2}\left(\frac{i\omega}{U}\right)\int_{x_a}^{x_b}\tilde{\phi}(x_1)^2\exp\left(i\frac{\omega}{U}(x_1-x_b)\right)(x_1-x_b)\,dx_1\\
&\hspace{1cm}-\left(\frac{i\omega}{U}\right)\int_{x_a}^{x_b}\int_{x_a}^{x_1}\tilde{\phi}(x_1)\tilde{\phi}(x_2)\exp\left(i\frac{\omega}{U}(x_2-x_b)\right)\,dx_2\,dx_1\\
&\hspace{1cm}-\left(\frac{i\omega}{U}\right)^2\int_{x_a}^{x_b}\int_{x_a}^{x_1}\tilde{\phi}(x_1)\tilde{\phi}(x_2)\exp\left(i\frac{\omega}{U}(x_2-x_b)\right)(x_2-x_b)\,dx_2\,dx_1\\
&\hspace{1cm}+\left(\frac{i\omega}{U}\right)\int_{x_a}^{x_b}\int_{x_a}^{x_1}\tilde{\phi}(x_1)\tilde{\phi}(x_2)\exp\left(i\frac{\omega}{U}(x_2-x_1)\right)\,dx_2\,dx_1\\
&\hspace{1cm}+\left(\frac{i\omega}{U}\right)^2\int_{x_a}^{x_b}\int_{x_a}^{x_1}\tilde{\phi}(x_1)\tilde{\phi}(x_2)\exp\left(i\frac{\omega}{U}(x_2-x_1)\right)(x_2-x_1)\,dx_2\,dx_1.
\end{aligned}
\end{equation}
Now we add up all the second order contributions using Eqns. \eqref{eqn:1st_term} \eqref{eqn:final_1} \eqref{eqn:final_2} \eqref{eqn:final_3} \eqref{eqn:final_4}. The majority of terms cancel out and we are left with
\begin{equation}
\begin{aligned}
\dot{P}_i\,\simeq&\,n_\infty m_i\dot{U}\frac{\psi^2}{U^4}\bigg[-3\int_{x_a}^{x_b}\tilde{\phi}(x)^2\,dx+4\left(\frac{i\omega}{U}\right)\int_{x_a}^{x_b}\int_{x_a}^{x_1}\tilde{\phi}(x_1)\tilde{\phi}(x_2)\exp\left(i\frac{\omega}{U}(x_2-x_1)\right)\,dx_2\,dx_1\\
&+\left(\frac{i\omega}{U}\right)^2\int_{x_a}^{x_b}\int_{x_a}^{x_1}\tilde{\phi}(x_1)\tilde{\phi}(x_2)\exp\left(i\frac{\omega}{U}(x_2-x_1)\right)(x_2-x_1)\,dx_2\,dx_1\bigg].
\end{aligned}
\end{equation}
The expansion of $\dot{P}_e$ in $\psi/U^2$ is trivial
\begin{equation}
\begin{aligned}
\dot{P}_e\,&=\,-m_e\dot{U}n_\infty\int_{x_a}^{x_b}h(\sqrt{\phi(x)})+1-\frac{\abs{U}}{\sqrt{U^2-2\phi(x)}}\,dx\\
&\simeq\,\,-m_e\dot{U}n_\infty\int_{x_a}^{x_b}h(\sqrt{\phi(x)})-\frac{\psi}{U^2}\tilde{\phi}(x)\,dx.
\end{aligned}
\end{equation}
We introduce the constant $I_0$ and a new function $I$
\begin{eqnarray}
&&I_0=\int_{x_a}^{x_b}\tilde{\phi}(x)^2\,dx, \\
&&I(\frac{\omega}{U})=\int_{x_a}^{x_b}\int_{x_a}^{y}\tilde{\phi}(x)\tilde{\phi}(y)\exp(i\frac{\omega(x-y)}{U})\,dx\,dy.
\end{eqnarray}
We thus get the final leading order expansion of $\dot{P}_i/\dot{P}_e$ as it appears in Eqn. \eqref{eqn:expansion}
\begin{equation}
\frac{\dot{P}_i}{\dot{P}_e}(\omega,U,\phi)
\,\simeq\,-\frac{m_i}{m_e}\frac{\psi^2}{U^4}\frac{4i\displaystyle \frac{\omega}{U}I(\displaystyle \frac{\omega}{U})+i\displaystyle \frac{\omega^2}{U^2}I'(\displaystyle \frac{\omega}{U})-3I_0}{\displaystyle \int_{x_a}^{x_b}h(\sqrt{\phi(x)})-\displaystyle \frac{\phi(x)}{U^2}\,dx}.
\end{equation}
\bibliographystyle{unsrt}
\bibliography{Hole_instability.bib}
\end{document}